\documentclass[sigconf]{acmart}

\usepackage[utf8]{inputenc}
\usepackage{subcaption}
\usepackage{mathtools} 
\usepackage{fancyvrb}
\usepackage{listings}
\usepackage{enumitem}
\usepackage{textcomp}
\usepackage{makecell}
\usepackage[colorinlistoftodos]{todonotes}
\usepackage{csquotes}

\usepackage[toc,page]{appendix}

\usepackage{amssymb}

\AtBeginDocument{%
  }

\copyrightyear{2026}
\acmYear{2026}
\setcopyright{cc}
\setcctype{by-nc-nd}
\acmConference[UIST '26]{The 39th Annual ACM Symposium on User Interface Software and Technology}{November 02--05, 2026}{Detroit, MI, USA}
\acmBooktitle{The 39th Annual ACM Symposium on User Interface Software and Technology (UIST '26), November 02--05, 2026, Detroit, MI, USA}
\acmDOI{10.1145/3830398.3830574}
\acmISBN{979-8-4007-2856-3/2026/11}

\begin{document}

% \title[Decompose-Demonstrate-Simulate]{Decompose-Demonstrate-Simulate: Bring Agency and Social Feedback Back into AI-assisted Style Development for Digital Artists}
\title[Analyze-Experiment-Resituate]{From Style Replication to Style Exploration: Enabling Art Style Exploration with Analyze-Experiment-Resituate Framework}
% \title{AER: Enabling Art Style Exploration with Analyze-Explore-Resituate Framework for AI-assisted Creativity Support}

\author{Wen-Fan Wang}
\authornote{Both authors contributed equally as first author.} % First author note
\email{vann@cmlab.csie.ntu.edu.tw}
\orcid{0009-0001-1050-1170}
\affiliation{%
  \institution{National Taiwan University}
  \city{Taipei}
  \country{Taiwan}
}

\author{TsaiHsuan Lin}
\authornotemark[1] % Share the first author note
\email{r13725039@cmlab.csie.ntu.edu.tw}
\orcid{0009-0001-7153-8081}
\affiliation{%
  \institution{National Taiwan University}
  \city{Taipei}
  \country{Taiwan}
}

\author{Chi-Lan Yang}
\email{chilanyang@slis.tsukuba.ac.jp}
\orcid{0000-0003-0603-2807}
\affiliation{%
  \institution{University of Tsukuba}
  \city{Ibaraki}
  \country{Japan}
}

\author{An-Ru Cheng}
\email{r14725059@cmlab.csie.ntu.edu.tw}
\orcid{0009-0005-6491-9969}
\affiliation{%
  \institution{National Taiwan University}
  \city{Taipei}
  \country{Taiwan}
}

\author{Bing-Yu Chen}
\email{robin@ntu.edu.tw}
\orcid{0000-0003-0169-7682}
\affiliation{%
  \institution{National Taiwan University}
  \city{Taipei}
  \country{Taiwan}
  }

\renewcommand{\shortauthors}{Wang Lin Yang Cheng Chen}

%% Generate the CCS concept using the tool at http://dl.acm.org/ccs.cfm
%===copy and replace the code generated from the website directly!===%


\begin{abstract}
% CHI-LAN
% Artistic style is a signature of professional digital artists and takes time to develop through repeated experimentation. Generative AI models can reproduce artistic styles with high fidelity. However, current generative AI tools do not allow artists to explore new styles, which could lead to a lack of stylistic diversity if they are only used to replicate dominant styles. To maximize GenAI's support for digital artists in exploring new styles, this study proposed an Analyze-Experiment-Resituate (AER) framework based on findings from interviews with 10 professional digital artists. We implemented this AER framework and evaluated its impact on 16 artists. We found that the AER framework enhanced artists' creative self-efficacy and confidence in pursuing new style directions, compared to using a direct style-transfer method. A further 2-week field study with 4 artists revealed that XXX. We discuss the opportunities and challenges of integrating the AER framework into the AI-assisted style exploration that balances artists' agency while co-exploring with generative AI.

Art style is a signature of professional digital artists that develops through repeated experimentation, reflection, and adaptation. While generative AI (GenAI) can reproduce styles with high fidelity, current tools provide limited support for exploring new stylistic directions and may encourage style replication over exploration.
To address this gap, we propose \textit{Analyze-Experiment-Resituate (AER)}, a framework for AI-assisted style exploration derived from interviews with 10 professional digital artists. 
Rather than prioritizing visually appealing outputs alone, AER supports three core practices of style exploration, including interpreting references, trying out stylistic possibilities, and reflecting on how emerging styles may be received. Specifically, AER enabled artists to (1) analyze artworks into interpretable stylistic elements, (2) have controllable experimentation guided by their own choices, and (3) resituate emerging styles through simulated social perspectives.
We implemented AER in a prototype system and evaluated it in a controlled study with 16 artists. Compared with a direct style-transfer workflow, AER increased artists' agency and reflection as they pursued new stylistic directions. A two-week field study with four artists revealed how the AER framework influenced daily style exploration, such as reflection, experimentation, and stylistic decision-making at each stage. We discuss opportunities and challenges in designing AI-assisted style-exploration workflows, and outline implications for future artistic support tools.

\end{abstract}



\begin{teaserfigure}
  \includegraphics[width=\textwidth]{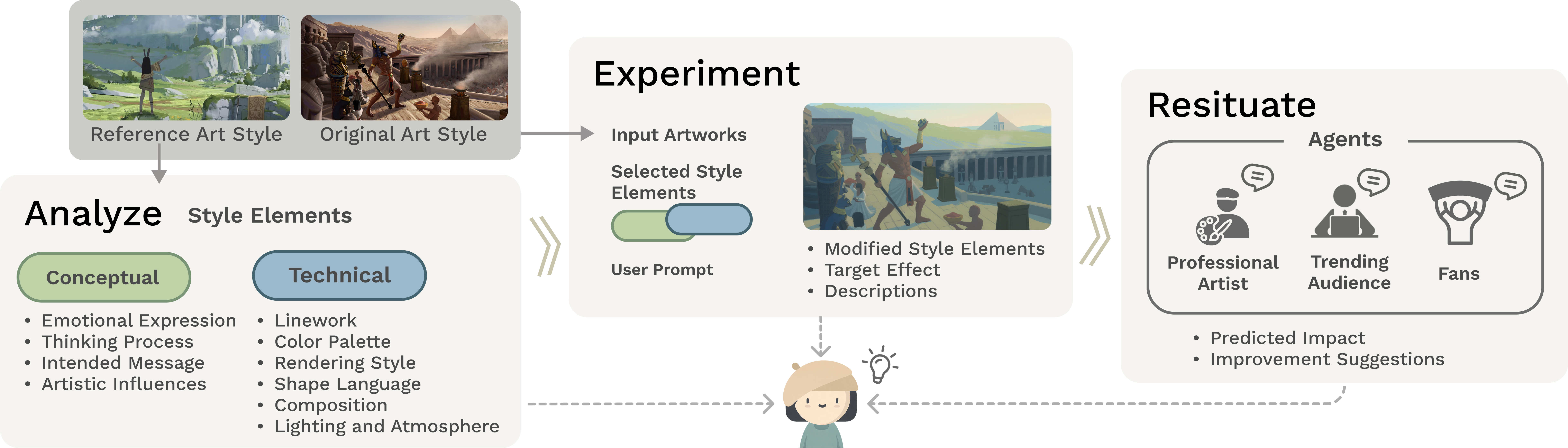}
  % \caption{The AER framework. The 
  % Analyze stage breaks down the reference artwork into style elements. The Experiment stage generates style directions based on inputs, and the Resituate stage provides agent feedback.}
  \caption{The Analyze-Experiment-Resituate (AER) framework. Existing GenAI tools encourage style replication over exploration. To enable professional artists to explore a new art style with GenAI, we developed the AER framework based on interviews with 10 professional digital artists. The AER framework allows artists to analyze both the conceptual and technical elements of a reference artwork during the \textit{Analyze} stage. In the \textit{Experiment} stage, it enables artists to experiment with new styles by employing prompt engineering and selecting pre-decomposed style elements. Additionally, in the \textit{Resituate} stage, it provides simulated feedback from role-based agents, including professional artists, audiences, and fans. This feedback helps artists reassess their developing styles from multiple perspectives.}  
  \Description{
The AER framework is a multi-stage process, which can be broken down into three main parts: Analyze, Experiment, and Resituate. The Analyze section begins with the reference artwork. On the left side, the artwork is broken down into style elements, which are further divided into two categories: conceptual (including emotional expression, thinking process, intended message, and artistic influences) and technical (including linework, color palette, rendering style, shape language, composition, and lighting and atmosphere). The Experiment section takes multiple inputs, including the original artwork, the reference artwork, selected style elements, and user prompts. On the right is the output image, along with its modified style elements, a target effect, and descriptions. The Resituate stage uses three personas: a professional artist, a trending audience, and fans. These personas review the generated image and provide feedback, which includes predicted impact and improvement suggestions.
}
  \label{fig:hero image}
\end{teaserfigure}


\maketitle

\section{Introduction}
\label{sec:introduction}

% Art style is a central part of professional visual creation. 
Art style is the soul of professional visual creation. For digital artists like illustrators and concept artists, style functions not only as a recognizable visual signature, but also a way of expressing artistic identity~\cite{abuhamdeh2015artistic}, emotional intent~\cite{guo2024emotional}, and creative thinking through recurring visual and conceptual choices~\cite{loomis1984cognitive, schapiro1994theory}. 
% At the same time, style is shaped by social and professional context, as artists develop and adapt it in relation to peers, clients, and audiences~\cite{becker2023art, bourdieu1969intellectual}.
In artists' creative practice and long-term development, style exploration is therefore an iterative process in which they study references, experiment with techniques and stylistic directions, reflect on their creative decisions, and incorporate feedback from peers~\cite{gelade2002creative, hu2018creative, kang2019art, schapiro1994theory}. 
% Through this process, artists may push beyond established habits, pursue new creative possibilities, or adapt their style to evolving project, market, and professional demands~\cite{porquet2025copying, williford2023exploring, skaggs2022trend}.

In the digital era, the style exploration process has been accelerated by tools that expand access to references and enable faster style experimentation. For instance, digital painting tools support efficient iteration with brushes, palettes, and rendering techniques~\cite{annum2014digital}, while online platforms expose artists to a much broader range of visual material and creative influences~\cite{wang2025aideation}. Generative AI (GenAI) further extends these possibilities by enabling rapid ideation~\cite{wang2025aideation, choi2024creativeconnect, yan2023xcreation}, image refinement~\cite{wang2025gentune, peng2024designprompt}, and exploration of visual alternatives~\cite{son2024genquery, wang2025aideation}. However, prior research has focused on supporting ideation or improving the speed and quality of image generation, while the design space for supporting style exploration remains underexplored. 

In much of AI research, style is operationalized as a set of aesthetic features that can be copied, transferred, or reproduced across images~\cite{porquet2025copying}. This view differs from how artists experience style in practice: not as a fixed visual property, but as an evolving and situated aspect of creative work~\cite{leitch2025unlimited} that requires substantial time, reflection, and iterative exploration. Current GenAI tools also tend to prioritize stylistic outputs over artists' interpretation and decision-making, reducing their sense of control and authorship~\cite{kawakami2024impact, jiang2023ai}. As a result, while there are many GenAI-based tools for creativity support, existing AI support offers limited value for style exploration. 
% To date, little research has examined how AI systems might support style exploration while preserving artists' creative agency.

% 有這個應該就不用寫最後的Contribution了？
Therefore, this work employed three studies to investigate the design space of GenAI-supported style exploration for professional illustrators and concept artists. We focused on illustrators and concept artists because they must continuously adapt their style to the needs of fans, audiences, and clients while maintaining a distinct artistic voice. First, we started by understanding artists' current practices and challenges when developing a new style. We asked, \textbf{\textit{what motivations, strategies, and challenges influence professional illustrators' and concept artists' style exploration practices?} (RQ1)} In the formative study, we examined how these artists explore and evolve their personal art style. We then developed an interaction framework, \textit{Analyze-Experiment-Resituate} (AER)\footnote{https://github.com/vannpacks/Analyze-Experiment-Resituate-An-AI-assisted-Art-Style-Exploration-Framework}, to inform the design of GenAI for style exploration based on the interview findings. Next, we conducted a within-subjects experiment with 16 professional illustrators and concept artists to examine how this AER framework influences style exploration compared with a direct AI-generated style transfer workflow. In the controlled experiment study, we answered, \textbf{\textit{how does the proposed AER framework influence artists' agency and reflection during style exploration?} (RQ2)}
Finally, we implemented this framework on a web platform and evaluated its influence on artists' style exploration processes and stylistic decision-making in their daily practice. 
We conducted a two-week field study with two professional illustrators and two professional concept artists, aiming to answer: \textbf{\textit{how do artists engage with the AER framework over time during style exploration?} (RQ3)}

Our contributions include an empirical account of visual artists' practices in exploring styles (formative study), an interaction framework to guide the design of creativity-support tools specifically for AI-assisted style exploration (controlled experiment study and field study), and a demonstration of how GenAI can be designed to enable visual artists to analyze, experiment, and reflect during their style exploration without compromising their creative agency.
Together, these contributions shift the role of GenAI in creative practice from merely producing \textit{stylistic outputs} as part of a supply chain toward supporting an artist-led \textit{process} in style exploration, enabling more open interpretation and reflection.

\section{Related Work}%artist definition of style 
% \subsection{What Does Style and Style Exploration Mean to Artists}
\subsection{Process Matters in Style Exploration}
For digital artists, style serves as both a recognizable signature and a marker of creative identity~\cite{panofsky1995three, ross2003style}. It facilitates recognition by allowing audiences and clients to identify an artist's work through recurring attributes such as color palettes, textures, strokes, and compositional patterns~\cite{pandey2024understanding, van2015toward}. 
% Beyond these surface features, 
Style embodies an artist's thinking process, personal expression, influences, and communicative intent~\cite{gombrich2018expression, gryglewski2020art, kemp2021artistic}. Furthermore, style also has a social dimension, connecting artists with cultural lines and communities while distinguishing them in professional contexts~\cite{sherman2017art, fischer1961art}. Based on these characteristics, this study defines \textit{style} as the visual, conceptual, and social signature of an artist.

%current understanding in the art field about style epxloration
The process of artists exploring style is defined as a continuous, recurring cycle, in which artists engage in a dynamic dialog between reference work, the tools and creative medium~\cite{gray2016visualizing}. It is a non-linear longitudinal process that often spans years, during which artists continuously adapt their style in the actual making of the art. This exploratory process involves the analysis of prior works and the reflection on feedback to iterate the style~\cite{porquet2025copying, leitch2025unlimited, jiang2023ai}.
%what motivates us to do formative study(no work on examining how artist explore style)
% Navigating the exploration of style is complex, but cultivating a unique style can serve as a marketable asset essential to an artist's livelihood and career~\cite{porquet2025copying}. However, despite the importance of style, a gap exists between the understanding of style exploration process and current research focus. 

Within the field of digital arts, existing studies center on the artifact-centric process of artwork production, which often focuses on the workflow of creating a specific work~\cite{wan2024breaking, sutton2013painter}. 
However, the digital support available to artists to explore new styles remains under-examined.
% Consequently, the process of exploring style remains under-examined. 
Our work addresses this gap by empirically examining digital artists' practices in style exploration, shifting the focus of creativity-support tools from artifact production to the process of style evolution.

% \subsection{What is Style and Style Transfer in GenAI-based Tools}
\subsection{The Invisible Process and Outcome-centered Style Transfer in GenAI-based Tools}
%how style was defined or simplified in tech field
Computational research has largely treated style as a transferable set of visual attributes, defined as a set of computationally reproducible factors, such as color and texture~\cite{ritchie2011d, leitch2025unlimited, hu2020aesthetic}. Treating style as a quantifiable set of visual features lays the computational foundation for style transfer research. 

%style transfer definition from tech viewpoints
Conceptually, style transfer involves extracting stylistic components from a reference image and applying them to other images. Early milestones such as Hertzman's Image Analogies \cite{hertzmann2023image}, framed style transfer as learning transformations from paired examples. This is followed by Gatys et al.'s Neural Style ~\cite{gatys2016image}, which disentangles content and style using CNNs, and Johnson et al.'s feed-forward networks for real-time performance~\cite{johnson2016perceptual}. Subsequent advances included GAN-based models for paired and unpaired translation, and diffusion models that made text-to-style generation broadly accessible, with ControlNet preserving structure fidelity while altering style~\cite{zhang2023adding}. More recent techniques emphasize lightweight personalization by generating stylized results from a single reference image and prompt~\cite{li2024styletokenizer}, or by combining a style reference with a separate content or structured input for finer control~\cite{batifol2025flux}. 
%what is missing or problematic in the current ai tools 
%problem1: treating style as ouput neglect creation process
% neglecting process leads to style homogenization and unable to support  reflection
However, these computational approaches frame style primarily as reproduction of output, ignoring the potential creative process involved when humans interact with these AI-infused systems. 

%style homogenization and unable to support reflection
This output-centric orientation can hide an artist's true intention and neglect the reflective practice that artists engage in to reassess stylistic choices. Particularly when GenAI tools impose a strong, built-in aesthetic direction, they override original expression causing the outputs of diverse users to become increasingly standardized. This loss of unique visual identity can subsequently lead to homogenization of style~\cite{epstein2023art, anderson2024homogenization, lin2026artificial}. 

%problem2: ai tool viewing style as visual features and reducing agency 
Furthermore, most digital creativity-support systems reduce style to reproducible visual features, optimized for fidelity rather than supporting artistic practice~\cite{jing2019neural, hu2020aesthetic}. Porquet et al. further critique that current style transfer tools provide little value to artists and risk diminishing their agency~\cite{porquet2025copying, kumar2025human}. GenAI systems often abstract away explicit prompting, translating high-level concepts into opaque intermediate representations that make it harder for artists to understand why specific images are produced, ultimately depriving artists of artistic control and agency~\cite{bernaschina2025visual}.

%ai tool is different from how artist workaflow 
%therefore, current hci research try to intergrate ai into artist workflow
%but existing research focus on supporting artwork creation and not style exploration
These approaches fundamentally clash with how artists approach their work. Although recent HCI research has started to look into the integration of GenAI into artists' creative workflows, the research focus has been on ideation~\cite{wang2025aideation, son2024genquery, cai2023designaid}, generation~\cite{wang2025gentune, peng2024designprompt}, or on fine-grained artistic control~\cite{lin2025sketchflex}, rather than supporting artists in exploring a new style~\cite{muller2025genaichi}.

% Additionally, a distinct line of research inquiry has integrated AI-generated feedback to facilitate reflection~\cite{yen2017listen, zheng2025artmentor, duan2024generating, chen2024memovis}, an important process artists engage in as they explore a new style. However, these technological interventions typically treat reflection as a means to refine a specific design or artwork rather than to influence artists' practices or ways of doing.

Additionally, prior work has used AI-generated feedback to facilitate reflection in creative practice~\cite{yen2017listen, zheng2025artmentor, duan2024generating, chen2024memovis}. Because artmaking is socially situated, feedback from peers and audiences can shape artists’ decisions, motivation, and creative identity~\cite{gao2019communication, chung2022association, simpson2025infrastructures, kim2017mosaic}. Recent systems have simulated audience or user perspectives through personas to support creative refinement and design evaluation~\cite{hung2024simtube, choi2025proxona, park2025stealing}. For instance, writers can specify AI personas that mirror their target readership~\cite{benharrak2024writer}. In theater, \enquote{Audience Amplified} used virtual spectators to increase social presence and engagement~\cite{kim2024audience}. However, these interventions primarily support reflection on a specific artifact or experience, leaving underexplored how AI-mediated reflection and simulated social perspectives may shape artists' longer-term stylistic practices.

% While these works offer valuable insights into how AI-assisted creativity tools can be designed, the focus remains on the execution of artifacts rather than enabling artists in the journey of stylistic exploration, a process of experimentation and reflection.
To shift the focus from outcome-centered style reproduction toward process-centered style exploration, we proposed a process-centered framework based on interviews with professional digital artists' style exploration practices and evaluated it to understand how to scaffold stylistic exploration with GenAI while maintaining the artist's creative agency.

\section{Method}
Our research employed a mixed-methods approach, including semi-structured interviews, questionnaires, and log data, to comprehensively understand artists' current practices when exploring new art styles, their perspectives on involving GenAI in this process, and how our proposed AI-assisted workflow influenced their style exploration. 
% We recruited digital artists, including concept artists and illustrators, who were from East Asian cultural contexts. Detailed participant information, including their participation in each study, is provided in Appendix ~\ref{Appendix_Participant_Demographic}. 
All studies received approval from the ethical review board of the authors' institution.

The research contains three parts of studies: (1) Formative Study (RQ1), (2) Controlled Experiment Study (RQ2), and (3) Field Study (RQ3). 
% In Formative Study, we focused on exploring the current practices of digital artists in developing their styles and identified design implications based on their needs. In the Controlled Experiment Study, we explored and evaluated how digital artists integrate the AI-assisted workflow under lab settings, which was designed based on the findings from the Formative Study, into their process of exploring a new style. Lastly, we conducted a two-week-long Field Study to investigate how the proposed framework influences artists in the long term.
The interview data were analyzed using thematic analysis \cite{braun2006using}. All interviews were transcribed and summarized. Two authors conducted the initial coding, one of whom had prior professional experience as a concept designer, and the resulting themes were iteratively discussed and refined within the research team until consensus was achieved.
% The interview data were analyzed using Reflexive Thematic Analysis (RTA)\cite{braun2019reflecting}. All interviews were transcribed and summarized. Two authors conducted initial coding, and final themes were then discussed and refined through group discussion to achieve consensus.

\section{Formative Study: Artists' Style Development: Practices, Challenges, and Design Goals}

% The Formative Study aims to answer RQ1, which includes (1) exploring current practices and motivations in artists' style development processes, and (2) identifying opportunities for GenAI to assist in these processes. 

\subsection{Participants}
We recruited 10 professional digital artists (ages 23-38; 6 male, 4 female), including 4 concept artists and 6 illustrators (with 4-15+ years of experience, mean = 9), who balance personal expression with client and viewer demands, making them well-suited for studying style development.
Participants were recruited through artist community connections and personal referrals by email to request collaboration. All the participants were based in East Asia.
Detailed participant information, including their participation in each study, is provided in Appendix ~\ref{Appendix_Participant_Demographic}. 
Participants were compensated 25 \$USD for a 1.5-hour interview study.

\subsection{Study Procedure}
First, we informed participants about their rights and obtained their consent to proceed with the interview.
The interview was conducted via Google Meet.
% to understand how artists develop their personal artistic styles. 
We asked participants to prepare examples of their personal work and project experiences beforehand, particularly focusing on their artistic style development and adjustment processes, including any artworks that could demonstrate their style development journey. Each interview lasted 1-1.5 hours.
% interview questions can be found in Appendix B.

\subsection{Findings}

\subsubsection{Practices and Challenges in Style Development}
% The development of artistic style is a complex process that involves imitation, learning, and the incorporation of personal preferences and habits over time.

The development of style is a complex process that requires substantial investment in time, research, deconstruction, and experimentation. It is not only an individual pursuit and exploration but is also influenced by social factors. 
% Across participants, we observed a recurring pattern in how they approached style exploration and development.

All artists mentioned that personal preferences drive their exploration. They began by collecting and exploring reference artworks by other artists, whether through actively searching or passively browsing. When artists find artworks they admire, they want to imitate and learn from them. As one participant explained, \enquote{\textit{I saw that artist's style and thought it was so cool, and the way it expressed using the texture made me want to try}} (P4).

% As one artist explained, "\textit{I just keep scrolling on Pinterest, [...], when I see artworks that I think are helpful to me, I will click into and then collect them}" (P2). 

A common next step was to carefully study other artists' works and break down their techniques from styles. We identified the aspects that the artists focus on, including \textit{composition, linework, color, lighting, rendering style, and shape language}. As one artist described, \enquote{\textit{I will find an artwork that I like, then I will break down its color use, techniques, shapes, and everything}} (P9). This analytical process goes beyond surface imitation. Instead, it emphasized understanding the reasoning behind stylistic choices. As P6 reflected, \enquote{\textit{I need to understand why others create that way, and what results that approach produces}}. In this sense, decomposition involved not only learning the \enquote{how} but also uncovering the \enquote{why} of artistic decision-making in developing a certain style. Finally, artists engaged in experimentation and adaptation, which sometimes involved conscious self-improvement or combining their preferences and habits. As P10 shared, \enquote{\textit{I have tried to imitate some artists' works... then integrate that into my personal drawing style.}} With repeated experimentation and considerable effort, artists incorporated these insights into their own evolving style practices.

% "\textit{I started thinking about what I learned from Japanese illustrations, [...], so I brought this line-focused habit into my Western-style creations}" (P7). 

% However, internal drives alone do not shape artistic style development. 
Additionally, social factors, including feedback from clients, viewers, and peers, significantly influence both the direction of decision and the details of style development. As P10 shared,  \enquote{\textit{Fans kept saying they loved my bold, saturated color strokes, so I leaned into that and made it a big part of my own style.}} At the same time, frustrations arising from not fully meeting client expectations or resonating with viewers may also lead artists to reflect on their personal style. 
% As P7 recalled, “\textit{At that time I worked on that [certain style] for a long time, but when I tried to transition my style, the clients kept asking for corrections. It was painful for about half a year.}” 
Advice from other professional artists could also provide essential reference points for refinement. For example, P8 noted, \enquote{\textit{When seeing the low engagement on social media for my works, I started seeking artist friends' suggestions [for adjusting my style].}} These examples show that style development is not solely an individual process but also a social one, where artists rely on social feedback to assist them in reflecting on their stylistic choices. 
% These show the ongoing importance of social feedback in style development, which helps reflective thinking on artists' style decisions.

% Challenges (併進去): 很花時間，投注成本 ，對於一些artist來說風險很大
% Artists sometimes find that their outcomes do not fully meet client expectations or resonate with viewers, which can lead to frustration. As P7 recalled, “\textit{At that time I worked on it [certain style] for a long time, but kept getting rejected by the clients when I transitioned my style, it was painful for about half a year.}”

\subsubsection{Challenges in Using GenAI in Style Development}

Interview findings showed that while some artists use AI for ideation and generating materials in their creative practices, none of our participants reported using GenAI or style transfer for style exploration. 
% We identified key challenges our participants faced when being asked about the reasons for not using GenAI in the style development process.

\paragraph{\textbf{Visually Appealing but Hard to See the Hidden Rationale}} First, AI images are difficult to analyze and fail to provide the information artists need for style development. Although AI outputs may appear visually appealing, they lack the process and underlying creation logic that typically guide human artworks (P1, P4, P5, P7, P8). From a technical perspective, AI images' illogical generation creates confusion for artists to try learning styles from them. As P8 explained, \enquote{\textit{[In AI images,] there are things you just can't make sense of, those things just confuse anyone trying to analyze it.}}

% {I feel like AI images have parts you can actually analyze and parts you really can't. The analyzable stuff might be like: Why does the color suddenly change right there? You can kind of look at those and think, oh, maybe that's why it looks good. But then there are things you just can't make sense of, like why is this area, which should be in shadow, suddenly so lit up? Those things just confuse anyone trying to analyze it.}". 
%% P8 原話: "我覺得AI圖有分有可以分析的跟不能分析的，可以分析的話也許就是，它的構圖為什麼是這個比例？為什麼在這個地方轉彎了？為什麼在這個地方突然顏色換了？這個也許可以去看，他這樣子為什麼好看。但是有些東西不能分析的，例如說為什麼這個地方突然亮了一塊？為什麼這個地方原本應該是在影子，可是它會變得這麼亮？那這個可能就會讓分析的人混淆。我在想，覺得它有應該的錯誤，它有可以分析學習的地方，也有沒有邏輯的地方。 "

% (暫時刪) 縮篇幅
%P4 highlighted a key difference between artworks made by humans and those created by AI, noting that "flaws" were often hidden from AI, but actually, they are helpful information to "show the artist's thought processes and problem-solving approaches".

%% While AI-generated content may look appealing, it lacks the creative logic, creator intent, and emotional depth that artists value when studying artistic works. 
%% As artists emphasized, "\textit{I cannot obtain deep information from AI-generated artworks because I cannot know the process of how the style is created.}" (P1, P7) 

% Beyond technical issues, AI images fall short in conveying deeper layers of meaning, including creators' intent and emotional depth, that come from intentional artistic decision-making, which is an important personal signature in one's style. 

Furthermore, AI-generated images fall short in conveying deeper meanings such as the creator's intent and emotional expression, which usually stem from intentional artistic choices. Artists described style exploration as a process that involves examining both the \textit{technical execution} and the \textit{conceptual thinking} behind other artists' works. However, since AI images remix countless unseen sources without clear attribution or rationale, they hide the artistic intent that is critical for learning and reflection (P2, P3, P10). This diminishes the value of AI-generated images as references.

%% 暫時刪
% As P8 further shared, "\textit{there are already far too many beautiful artworks nowadays, [...], what distinguishes meaningful art is the ability to communicate emotions and stories.}" 

%% 暫時刪
% Artists highlighted the importance of tracing inter-artist influences to understand stylistic development, such as how [An artist's work] might be inspired by [B artist]. Yet with AI, this becomes difficult; its outputs remix countless unseen sources without attribution, obscuring the conceptual artistic intent that is critical for learning and reflection (P3). 
% Artists view style development as a process that involves exploring and analyzing both the technical and conceptual aspects of other artists' works. However, AI-generated images do not facilitate this analysis of technical and conceptual aspects, which is essential for effective style development.
% Artists view style development as an exploration and analysis of both technical and conceptual aspects of others' artwork. However, AI-generated images could not enable them in analyzing the technical and conceptual aspects of the artwork, which is a crucial analytical process for style development..
% however, AI-generated images contribute little to this process.

\paragraph{\textbf{The Deprived Creative Agency}}
The second challenge concerns the loss of creative agency. As P1 noted, \enquote{\textit{GenAI tends to produce complete designs that create a false sense of completeness.}} Such completeness deprives artists of opportunities for intervention, modification, or further contemplation. Furthermore, artists worry that AI-generated references can undermine their creative style development and independent thinking. As P7 admitted, \enquote{\textit{If I check AI results while creating, my ideas get locked in too easily. I only use AI when clients have clear requests, then I can just work with that constraint. But for developing my own style, I almost never use AI.}} 
% This dependency threatens the core of artistic growth. 
% Several participants voiced concerns about the risks of over-reliance on AI, suggesting that it may undermine the process of developing personal habits, design ideas, and creative identities (P2, P5, P6). They believe this could restrict style development to the limitations of AI's training data, rather than supporting authentic self-development.

These challenges explain why current AI and style transfer models do not meet artists' needs for style development. While AI may excel in certain technical aspects (e.g., generating appealing images), it falls short of supporting the deeper, more intentional processes that artists require for meaningful style development and growth.

% %% Artist's Perspective to AI
% %% style transfer outputs offer limited reference value because they obscure creators' intent, emotion, and creative logic
% The artists suggested that although AI-generated content may possess surface-level aesthetic appeal, its difficulties in conveying creative intent, emotional depth, and learning opportunities significantly constrain its value as reference material for serious artistic style practice and development.

\subsection{Design Goals}
Based on the findings, we identified three design goals for integrating GenAI into artists' style exploration practices.
\begin{itemize}
% How do we integrate GenAI into the style development workflow?
    \item \textbf{DG 1: Keep Artistic Agency when Co-exploring with GenAI:} Empowering artists to reflect, critique, and steer the direction of style exploration, rather than passively consuming AI outputs.
    
    \item \textbf{DG 2: Make Style Exploration more Analytical and Explicit:} Help artists expand their interpretation of the \enquote{how} and \enquote{why} behind stylistic outputs by breaking down reference elements (technical aspect) and revealing underlying aesthetic principles (conceptual aspect).

    \item \textbf{DG 3: Integrate Social Perspectives into Individual Style Exploration:} AI can simulate potential feedback from representative audience groups, enabling artists to reflect on how their work may be interpreted and received by others.
    % \item \textbf{DG 3: Bring Social Aspect back into an Individualized Style Development Process:} While current AI tools often emphasize individual creativity \cite{choi2024creativeconnect,wang2025aideation, feng2023promptmagician}, our Formative Study showed that style development is inherently social (e.g., engaging with peers' work, gathering viewer feedback). AI can thus serve not only as a creative assistant but also as a means to simulate diverse social feedback. This could enable artists to infer social reception and support reflection on how a style communicates emotion, identity, or values. By doing this, artists can make informed decisions about exploring new styles without investing significant time and resources upfront.
    % AI can thus serve not only as a creative assistant but also as a means to simulate audience perception, predict social reception, and support reflection on how a style communicates emotion, identity, or values.

\end{itemize}

\section{Analyze, Experiment, and Resituate: A Framework for Creative-support Tools}

\begin{figure*}[ht]
  \centering
  \includegraphics[width=\textwidth]{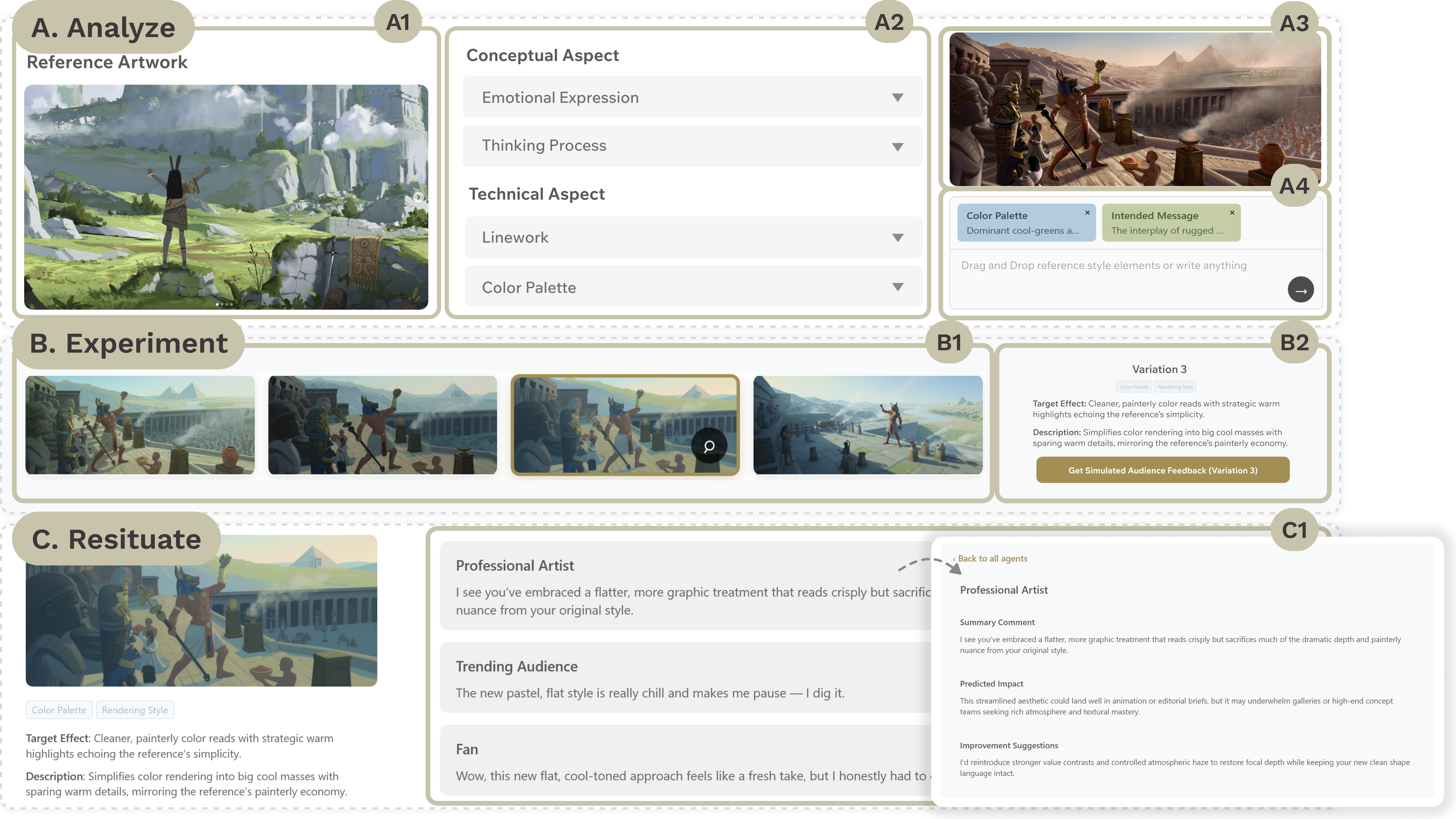}
  \caption{
  % The interface of [AER-embedded system] includes: (A) The \textit{Analyze} panel. Users upload a reference artwork, which serves as a style reference, and a personal artwork representing the current style. The reference artwork is broken down into technical and conceptual aspects, presented as various elements. The user can drag elements into the chatbox. (B) The \textit{Experiment} panel. The panel presents four style-tweaking directions based on user-selected elements. Once a variation is selected, a textual explanation that contains the target effect and description is displayed. (C) The \textit{Resituate} panel. Users can click “Get Simulated Audience Feedback” to simulate three responses based on the role of \textit{Professional Artist}, \textit{Trending Audience}, and \textit{Fan}. Each feedback consists of a summary, predicted impact, and suggestions for improvement. 
  The AER-embedded system consists of three panels: (A) \textit{Analyze}, which decomposes reference artwork into technical and conceptual elements; (B) \textit{Experiment}, which generates and explains stylistic variations based on selected elements; and (C) \textit{Resituate}, which provides simulated feedback from professional artist, trending audience, and fan perspectives.
  }
  \Description{The figure shows three stages of UI of the AER-embedded system, labeled Analyze, Experiment and Resituate. This Analyze section is composed of four main parts: Referenced Artwork at the top left, contains the source image which style will be analyzed and serves as the stylistic reference. To the right is the Conceptual and Technical Aspect Analysis, which features a series of expandable dropdown menus that break down the source image's style into elements. These are categorized under two headings: Conceptual Aspect and Technical Aspect. On the right side is the Target Tweaking Style which contains the area of the target image that will apply the new style. Under which is a drag and drop interactive panel that allows the user to specify which stylistic elements should be applied to the Target Tweaking Style. In the figure, two tags have been added: Color Palette and Intended Message. An arrow icon is within the chatbox for proceeding to the next stage. The Experiment section contains an image grid showcasing four versions of the generated images representing different style tweaking directions. At the right is the variation details, which displays details once a variation from the image grid is selected. The panel is titled Variation 3 and specifies the stylistic elements applied such as Color Palette and Rendering Style. The details of the variation provides a Target Effect description and Description of the rendering process. The Resituate section is the simulated audience feedback. This panel presents qualitative feedback from three roles to provide critiques, including Professional Artist noting “the new style is a flatter, more graphic treatment that reads crisply but sacrifices much of the dramatic depth and painterly nuance from your original style.” Trending Audience states “The new pastel, flat style is really chill and makes me pause — I dig it.” Fan comments, “Wow, this new flat, cool-toned approach feels like a fresh take, but I honestly had to do a double-take without the lush details I adore.” To the right of these comments, a separate sidebar shows the interface once a feedback is clicked upon. It provides Summary Comment, Predicted Impact and Improvement Suggestions from the Professional Artist feedback.
}
  \label{fig:system_ui}
\end{figure*}

Building on insights from Formative Study, we present the Analyze-Experiment-Resituate (AER) framework within an AI-assisted style exploration workflow. 
The AER framework aims to embed creative agency and social feedback into the generative process.
To examine its effectiveness, we developed a proof-of-concept system that operationalizes the AER workflow and enables empirical evaluation through user studies.

\subsection{Overview of the AER Framework}
The Analyze-Experiment-Resituate (AER) framework structures the AI-assisted style exploration process into three stages, addressing the design implications discovered in the formative study. This AER framework is designed to enable artists with an established style to explore new styles by integrating their existing style with others they wish to pursue.
\subsubsection{Analyze}
\textit{Analyze} refers to the process by which GenAI breaks down the technical and conceptual elements of an artwork. 
Grounded in the practices identified in our formative study, where artists dissect and reason through the artworks they admire, this stage, our system extracts both technical elements (e.g. composition, shape language, linework, lighting and atmosphere, color palette and rendering style) and conceptual elements (e.g., emotional expression, cognitive process, intended messages, and artistic influences). 
The \textit{Analyze} stage addresses the challenge of interpretability by surfacing specific layers of an artwork rather than simply presenting a style-copied piece. This approach aims to help artists understand how a style is constructed and how stylistic choices are made. The generated analyses are designed to prompt reflection during style exploration as provisional interpretive resources.

\subsubsection{Experiment}
\textit{Experiment} refers to the process by which GenAI presents various combinations for artists to experiment with different technical and conceptual elements while exploring a new style.
This stage enables artists to explore stylistic variations actively, allowing them to guide the style generation using decomposed elements from reference artworks in the \textit{Analyze} stage or their own thoughtful input.
At this stage, the system allows artists to specify and choose the technical or conceptual elements they want to incorporate into their artwork. The system then generates several interpretable variations, each accompanied by detailed explanations outlining what was changed, how it was applied, and the rationale behind each direction taken.
This balance between active control and diverse outputs enables artists to experiment in various directions while maintaining an understanding of the underlying stylistic logic, which is a critical challenge when using a style transfer model to directly apply one's style to another artwork.
Unlike directly copying one style to another, the \textit{Experiment} stage preserves artistic agency and encourages experimentation and authorship. Artists are not passive recipients of AI output but remain central in driving stylistic direction.

\subsubsection{Resituate}
\textit{Resituate} refers to the process in which the AI-assisted tool simulates the social feedback of various social roles that artists consider in their style exploration practices. Building on the findings from the formative study, which indicate that artists depend on peer critiques and viewer reactions to adjust their styles, this \textit{Resituate} stage presents social feedback articulated in the voices of key social roles that artists value. These roles, such as a fan, a professional artist, or a general audience, are grounded in our formative study results and supported by prior literature~\cite{seki2013finding, day2016effect}.
Each feedback role shared comments, predicted impacts, and suggestions for improvement from their perspective, aiming to allow artists to preview how different viewer groups might interpret or respond to a given style artists plan to explore.
% In doing so, \textit{Resituate} enables artists to assess possible social reactions before they commit significant time and effort to developing a new stylistic approach.

Together, these three stages show how GenAI can be integrated into professional artists' workflow of style exploration: decomposing references in the \textit{Analyze} stage, trying out variations in the \textit{Experiment} stage, and seeking feedback in the \textit{Resituate} stage.

\subsection{Implementing AER-embedded System}
\label{sec:AER-system}

\begin{figure*}[ht]
  \centering
  \includegraphics[width=\textwidth]{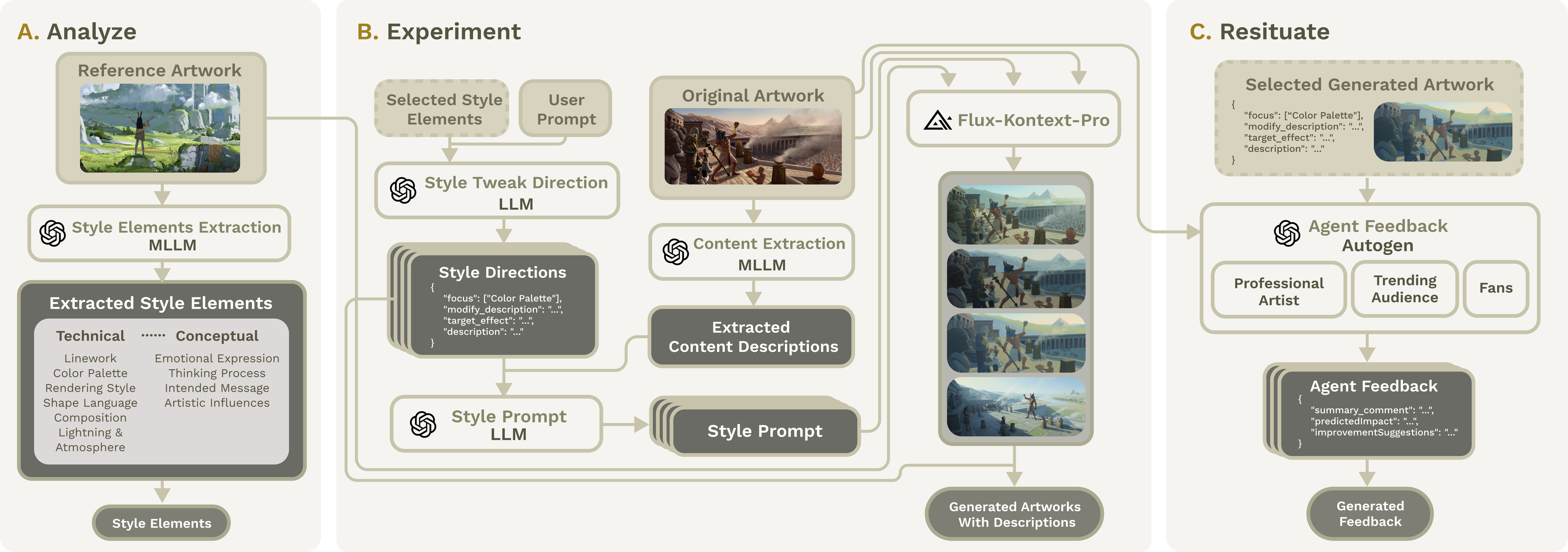}
  \caption{The system diagram for AER-embedded system. 
  %Pipeline for Analyze: Once a reference image is provided, Style Elements Extraction MLLM extracted technical and conceptual style elements from the reference artwork.  Pipeline for Experiment: The Style Tweak Direction LLM takes user selected style elements and user prompts as input to construct the style direction. Content Description Extraction MLLM extracts content descriptions from the user's original artwork. Style Prompt LLM take the style direction and the extracted content descriptions to construct the style prompt, which is passed to Flux-Kontext-Pro along with the original and reference artwork to generate artworks with descriptions. Pipeline for Resituate: if a user selects a generated artwork, Agent Feedback AutoGen framework takes both the selected artwork and the original artwork to produce agent feedback, and generates simulated feedback.
  }
  \Description{The image displays the system diagram, and is organized into three vertical sections labeled Analyze, Experiment, and Resituate, with arrows indicating the flow of data and processing steps. This initial Analyze stage begins with a Referenced Artwork, which is fed into a Style Elements Extraction MLLM. The MLLM produces extracted style elements with detailed descriptions and categorizes the extracted elements into Technical and Conceptual. This structured data is then passed on as the final Style Elements output of this stage. The second stage is Experiment. The Selected Style Elements, User Prompt is passed into Style Tweak Direction LLM, and it generates Style Direction. While Original Artwork is passed into Content Description Extraction MLLM, it produces Extracted Content Descriptions, which is a structured content summary of the original image. Both the Style Direction and Extracted Content Descriptions are then fed into a Style Prompt LLM that synthesizes Style Prompt that details the content and the desired stylistic changes. Style Prompt, along with referenced artwork and original artwork, is passed to the generative model Flux-Kontext-Pro, which produces the final visual output with a description. The final stage is Resituate, which begins with a Selected Generated Artwork, which is passed to the Feedback Autogen framework. The framework includes three distinct personas: Professional Artist, Trending Audience, and Fans. The framework generates structured Agent Feedback containing a summary, predicted impact, and improvement suggestions for the given role.
}
  \label{fig:system_diagram_horizontal}
\end{figure*}

% This section describes the system architecture, interaction flow, and design features that bring AER into practice.
Figure~\ref{fig:system_ui} shows the system's user interface, Figure~\ref{fig:system_diagram_horizontal} presents the system diagram, and the complete interaction workflow is provided in Appendix \ref{Appendix_workflow}.
To use this system, users have to upload two artworks: one that represents their current style (Fig.~\ref{fig:system_ui}-A3) and another that serves as a reference for learning or inspiration (Fig.~\ref{fig:system_ui}-A1). 
This mirrors common artistic practice, as seen in our formative study and prior literature, where stylistic elements are extracted from a reference and applied to one's own work~\cite{muller1982rubens}. Likewise, state-of-the-art style transfer methods typically pair a style reference with a content or structural reference~\cite{batifol2025flux}. 

\subsubsection{Analyze}
The system adopts GPT-5\footnote{GPT-5, https://platform.openai.com/docs/models/gpt-5}
as its primary multimodal large language model (MLLM). 
Prior studies have validated the capability of MLLMs to analyze stylistic elements in artworks~\cite{zheng2025artmentor}. 
The system analyzes the reference artwork through a dedicated prompt designed to identify its technical and conceptual aspects (Fig. ~\ref{fig:system_diagram_horizontal}-A).
% For each conceptual component, the system also identifies which technical elements contribute to it. 
These components are visualized as tags in the web-based interface (Fig.~\ref{fig:system_ui}-A2), which users can drag into the chatbox to guide subsequent generations (Fig.~\ref{fig:system_ui}-A4). For example, under the technical aspect \textit{Rendering Style}, the system may output: \enquote{\textit{Painterly, with broad, confident brush masses and visible strokes that suggest texture rather than detail.}} Under conceptual aspects, it might describe: \enquote{\textit{The sweeping vista and clear atmospheric ladder suggest openness, scale, and forward momentum. By arranging large directional planes and keeping the palette fresh and cool, the image nudges the viewer toward thoughts of exploration and possibility.}} This conceptual analysis is accompanied by three related technical tags: composition, color palette, and lighting \& atmosphere.

\subsubsection{Experiment}
After reviewing all analyzed aspects, the participant selects an emotional expression component, tagged with the rendering style and the color palette, and an additional rendering style component as input. A Style Tweak Direction LLM then combines the selected style elements with the user prompt to generate style-tweaking directions (Fig. ~\ref{fig:system_diagram_horizontal}-B). In parallel, a Content Description Extraction MLLM extracts content descriptions from the user's original artwork. A Style Prompt LLM subsequently integrates the style directions and extracted content descriptions to construct generation prompts. The system generates four style tweaking directions based on these selections, and constructs dedicated prompts for stylistic-controlled image generation.
We used Flux-Kontext-Pro\footnote{Flux-Kontext-Pro, https://bfl.ai/models/flux-kontext}
as the image generation model for its ability to integrate features from two images into one output. Each prompt combines the content of the original user image with a style modification instruction, ensuring that only the intended changes are applied while preserving other content.
Each variation is paired with a textual explanation (Fig.~\ref{fig:system_ui}-B1, B2). In this example, the first variation aimed to \enquote{\textit{create cleaner, painterly color reads with strategic warm highlights echoing the reference's simplicity}}, modifying the \textit{color palette} and \textit{rendering style}. The style change is explained as: \enquote{\textit{Simplifies color rendering into big cool masses with sparing warm details, mirroring the reference's painterly economy.}}

\subsubsection{Resituate}
Once the user identifies an image of interest, they click \enquote{Get Simulated Audience Feedback} to receive responses from the three aforementioned viewer roles. We use AutoGen~\cite{wu2024autogen} as the framework for generating agent feedback, with system prompts specifying the characteristics of the role (Fig. ~\ref{fig:system_diagram_horizontal}-C). The simulated professional artist focused on sharing constructive critique, the simulated trending audience on immediate social-media appeal, and the simulated fan on stylistic consistency and recognizability. The interface first displays summary comments from each agent (Fig.~\ref{fig:system_ui}-C1). For instance, the professional artist might respond: \enquote{\textit{I see you've embraced a flatter, more graphic treatment that reads crisply but sacrifices much of the dramatic depth and painterly nuance from your original style.}} Users can then select a specific agent's feedback to review more detailed outputs, including predicted impact: \enquote{\textit{This streamlined aesthetic could land well in animation or editorial briefs, but it may underwhelm galleries or high-end concept teams seeking rich atmosphere and textural mastery}}, and improvement suggestions such as \enquote{\textit{I'd reintroduce stronger value contrasts and controlled atmospheric haze to restore focal depth while keeping your new clean shape language intact.}} Then, after viewing the feedback, the user can type into the chatbox to begin a new round of exploration.
\section{Controlled Experiment Study: Exploring and Evaluating AER Framework}

% Controlled Experiment Study evaluates the effectiveness of the AER-embedded framework while supporting professional artists' style development practices. 
We conducted a within-subjects study with 16 professional digital artists to answer: \textbf{(RQ2) How does the proposed AER framework influence artists' agency and reflection during style exploration?}
% \textbf{(RQ2) how do different AI interaction structures shape artists' agency, reflection, and stylistic decision-making during style exploration? }

\subsection{Experiment Design}
We compared the proposed AER-embedded system with a direct style transfer baseline, as it is one of the AI-assisted tools most closely associated with stylistic creativity and has been widely used in both research and practice as a means of \enquote{copy styles} \cite{porquet2025copying}. We implemented the baseline with a similar interface to reflect common direct style transfer workflows.
% in order to know how our proposed framework different from one of the most accessible approaches for artists to use GenAI for style exploration.

% Style transfer is one of the AI-assisted tools most closely associated with stylistic creativity and has been widely used in both research and practice as a means of “copy styles” \cite{porquet2025copying}.

% To compare AER framework with one of the accessible approaches to using GenAI for style exploration, we implemented a non-AER-embedded system with 
\label{sec:non-AER-system}
The core interaction of the non-AER-embedded (direct style transfer) system was adapted from MidJourney\footnote{MidJourney, https://www.midjourney.com/home}, a widely used GenAI service. It included (1) uploading style reference artworks, (2) uploading personal artworks as content references, (3) a text input area for prompting, and (4) a section showing generated results. The backend used Flux-Kontext-Pro, the same image generation model as the AER-embedded system. It also utilized the same prompt structure that merged content descriptions from the user's artwork with style descriptions from the reference artwork. The user's prompt guided how these style elements were integrated, producing outputs that reflected this combination. The interface of the non-AER-embedded system is shown in Appendix ~\ref{Appendix_Baseline}.

During the study, participants used both systems to explore new styles with two types of artworks prepared in advance: (1) their own works or projects representing their current style, and (2) reference artworks they admired and wished to explore stylistically. 

The order of conditions was counterbalanced across participants. In both conditions, participants were asked to explore potential stylistic directions, with the resulting outputs serving as inspirational references rather than final artworks. Participants were also encouraged to think aloud when using the system. Outcomes could include AI-simulated illustrations, insights gained during exploration, and possible next steps for their style development.

\subsection{Participants}
We recruited 16 professional digital artists (age 23-38; 9 male, 7 female), including 6 concept artists and 10 illustrators (4-15 years of experience, mean = 7.25). Seven participants (P1, P2, P3, P4, P5, P7, P9) had participated in the Formative Study and were re-invited via email. The remaining participants were recruited using the same approach used in the Formative Study. 
%Detailed participant information is provided in Appendix ~\ref{Appendix_Participant_Demographic}.
All participants received 35 ~\$USD compensation for a 2-hour study. 

\subsection{Procedure}
The study began with a 10-minute briefing. Participants were informed about the study purpose, their rights, and provided consent before proceeding. 
For each condition, the session included a 10-minute tutorial, 30-40 minutes of style exploration, and a 5-minute post-task questionnaire after participants used each system. The study concluded with a 30-minute semi-structured interview. 

\subsection{Measurement}
% Questionnaire
% 要寫一下怎麼做study analysis
%% Behavior Log
% We adopted a mixed-method approach, including the self-reported questionnaire and the semi-structured interview. The measures collectively provide both quantitative and qualitative insights into how the framework influenced artists' style exploration.
%% evaluating the study with two complementary measures

%% We adopted a self-report approach, consistent with prior HCI and creativity research~\cite{lubos2024llm,satyanarayan2019critical,palani2022don,son2024genquery}, and analyzed the data using the Wilcoxon signed-rank test~\cite{woolson2005wilcoxon}, 

\begin{enumerate}
    \item \textbf {Self-reported Post-task Questionnaire}: Participants completed three validated questionnaires: Agency \cite{tapal2017sense}, Creative Self-Efficacy \cite{chen2001validation}, and Technology-Supported Reflection Inventory (TSRI) \cite{bentvelzen2021development}. All items were rated on a 7-point Likert scale (1: Strongly Disagree, 7: Strongly Agree). We adapted the items to reflect style exploration; for instance, \enquote{\textit{I am in full control of the style decisions I make using this approach.}} 
    % Two items were removed from the Creative Self-Efficacy scale as they were not suitable for our study context. 
    All scales demonstrated high internal consistency: Agency ($\alpha = .86$), Creative Self-Efficacy ($\alpha = .90$), and TSRI ($\alpha = .91$), including its sub-constructs of Insight ($\alpha = .86$), Exploration ($\alpha = .82$), and Comparison ($\alpha = .78$). Detailed items are provided in Appendix ~\ref{Appendix_Questionnaire}. After confirming normality via Shapiro-Wilk tests, we analyzed the data using paired t-tests and calculated Cohen's d for effect sizes.

    \item \textbf {Semi-Structured Interview}: We conducted in-depth semi-structured interviews to triangulate the quantitative results and examine how artists engaged with the AER-embedded system compared to direct style transfer. 
    We focused on understanding participants' agency, creative process, sense of future direction in style development, and perspectives on AI-assisted approaches while using two systems.

\end{enumerate}

\subsection{Findings}
% In this section, we report results and findings from the controlled experiment study.

\begin{figure}[ht]
  \centering
  \includegraphics[width=\linewidth]{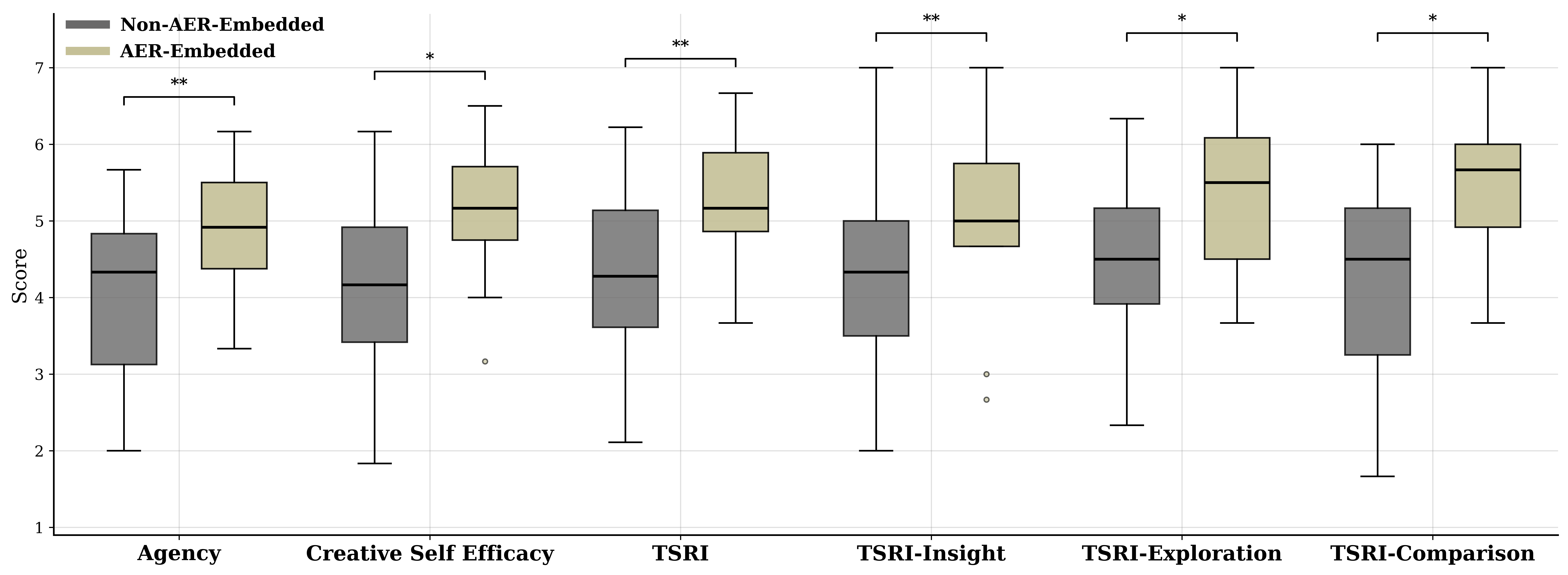}
  \caption{Questionnaire results from the within-subjects task. Participants rated Agency, Creative Self-Efficacy, and the Technology-Supported Reflection Inventory (TSRI), TSRI-Insight, TSRI-Exploration, TSRI-Comparison for both the Non-AER-Embedded and AER-Embedded system using a 7-point Likert. scale. *: p < .05 and **: p < .01.} 
% Participants rated Agency, Creative Self-Efficacy, and the Technology-Supported Reflection Inventory (TSRI), TSRI-Insight, TSRI-Exploration, TSRI-Comparison for both the Non-AER-Embedded and AER-Embedded system using a 7-point Likert. scale. *: p < .05 and **: p < .01.
  \Description{Boxplots comparing questionnaire scores across six evaluation metrics. The vertical axis shows scores from 1 to 7, and the horizontal axis lists six metrics: Agency, Creative Self-Efficacy, TSRI, TSRI-Insight, TSRI-Exploration, and TSRI-Comparison. Across all metrics, the AER-embedded condition shows significantly higher score distributions than the comparison condition.}
  \label{fig:result}
\end{figure}

\subsubsection{\textbf{AER-embedded System Increased Artists' Agency}}
%% Questionnaire: Agency
Agency in this context refers to the sense of truly driving the direction and outcome of style exploration.
Result of the paired samples \textit{t}-test showed that participants using the AER-embedded system reported significantly higher perceived agency compared to the direct style transfer (Figure~\ref{fig:result}; $t[15]=2.95, p = .010, d = 0.85$, mean difference = 0.88). 
Qualitative analysis of the interview further revealed three main reasons underlying this enhanced sense of agency, with many explicitly attributing their experience to the \textit{Analyze} and \textit{Experiment} mechanisms.

% \textit{Decision-making Influence (Traceability).} 
\paragraph{Traceability} 
Artists emphasized that agency is grounded in whether they could clearly see how their decisions influenced the output and whether the reasoning behind those decisions was traceable. P5 described, \enquote{\textit{After selecting analyzed elements, I can see how my choices change the results, and the side-by-side display of tweaking directions helps me better understand what I'm referencing.}} In contrast, the outputs from the non-AER-embedded system often felt opaque. As P14 noted, \enquote{\textit{[When looking at the output from the Non-AER-embedded system] I don't know how it adopted the artworks or what elements it actually used.}}

%% P2 "最有主導感應該是挑選的時候，挑個兩三次，真的有感覺它因為我挑選而產生變化，會感覺最有主導感。"
%% P16 "non-AER-embedded 生成結果上我「不知道它怎麼採用」、不知道它「採用了什麼」"

%% 期蘭comment:如果有人有提到這些decomposed element跟他們想的一致或是不一致，建議可以加上

\paragraph{Controllability}
% Artists also felt a stronger sense of ownership over stylistic outcomes while using the AER framework. 
% By making choices from decomposed elements, artists felt the results were authentically theirs instead of AI-determined. 
By making choices from analyzed elements in the  AER-embedded system, artists felt they were creating rather than receiving what AI had determined for them.
As P11 reflected, \enquote{\textit{This felt like I was truly creating, because I was making choices.}} While most artists favored the AER-embedded framework, a few found a different form of control in direct style transfer. For P16, the agency resides in the \enquote{\textit{deliberate process of formulating the prompt}}. While P16 acknowledged that AER allows better guidance of the creative direction afterward, the artist viewed the initial act of defining the prompt as also the expression of agency.

% UIST刪
% Likewise, P12 emphasized, “\textit{[Compared to direct style transfer, AER-embedded] lets me decide directions, rather than having AI assume it fully understands what [style] I want and directly creates the illustration [by copying the style from the reference artwork]. This [AER-embedded] feels more controllable.}”

% \paragraph{Alternative Views on Agency.} 
% 移來這裡

% “\textit{I chose from decomposed elements and integrated them into what I wanted. This felt like I was truly creating, because I was making choices.}”

\paragraph{Alignment with Artistic Identity}
The sense of agency was further supported when artists felt the AER framework aligned closely with their artistic identity.
% making it easier to direct their style exploration. 
% Participants described AER as an “art-literate” partner that supports deeper exploration and reflection, often likening the experience to conversing with someone who truly understands art. 
Participants described AER as a partner that supports deeper exploration and reflection, often likening the experience to conversing with someone who truly understands art.
For example, P1 noted that its analysis \enquote{\textit{spoke his language}}, lowering his guard, while P9 described it as \enquote{\textit{like chatting with another artist}} with clear and familiar logic.

%% UIST刪
% Beyond this sense of recognition, participants also emphasized its role in fostering genuine exploration and learning. P2 highlighted how AER helped her perceive the evolution of her own thinking and expand her stylistic vocabulary through textual descriptors such as “low saturation.”

% Taken together, although the qualitative results suggest that definitions of agency varied among participants, the AER-embedded framework consistently offered more opportunities to strengthen agency by making stylistic decisions traceable and ensuring that final outcomes felt authentically artist-driven rather than system-determined.

\subsubsection{\textbf{AER-embedded System Increased Reflection and Confidence in Future Style Development}}

%% Questionnaire result Self-Efficacy, TSRI
Participants using the AER-embedded system reported significantly higher levels of self-efficacy (Figure~\ref{fig:result}; $t[15]=2.90, p = .011, d = 0.88$, mean difference = 1.00) and technology-supported reflection (TSRI) (Figure~\ref{fig:result}; $t[15]=3.21, p = .006, d = 0.92$, mean difference = 0.96) compared to the non-AER-embedded style transfer. A closer look at the TSRI sub-scales revealed that AER framework strengthened artists' ability to gather \textit{Insight} ($t[15]=3.54, p = .003, d = 0.67$, mean difference = 0.85), to engage in \textit{Exploration} of style possibilities ($t[15]=2.50, p = .025, d = 0.78$, mean difference = 0.90), and to draw \textit{Comparisons} in a social sense ($t[15]=2.70, p = .016, d = 0.97$, mean difference = 1.12). 
% To better understand these findings, we examined how the different steps of the AER workflow shaped artists' creative processes.

%% UIST 刪 (Merger到後面)
% \paragraph{\textbf{How each step influences the creative process:}} 
% Each stage of the AER framework supported creativity in distinct yet complementary ways. \textit{Analyze} primarily stimulated divergent thinking and clarified stylistic possibilities. \textit{Demonstrate} then converged these insights into concrete visualizations that artists could selectively adopt. Finally, \textit{Simulate} provided reflective perspectives, functioning as feedback and suggestions to guide subsequent implementation.

% \paragraph{Analyze.} 
% Artists consistently described Analyze as surprisingly powerful, as it enabled them to better interpret the meaning behind references, extract stylistic elements, and apply them more effectively.
Interview results showed that \textit{Analyze} helped artists interpret the meaning of references and identify stylistic elements they had not previously articulated. P5 reflected, \enquote{\textit{I realized I liked the high contrast and vector-like color blocks. In the future, I can search for more styles based on this keyword.}} It also sparked new directions for exploration: \enquote{\textit{It's like an extra spark of interpretation... opening up more possible directions [for style exploration]}}(P19). While interacting with the stylistic elements in the \textit{Experiment} stage, artists further described the generated images as practical references that helped them clarify direction and anticipate challenges when trying out a style. As P11 noted, \enquote{\textit{[It] helps me realize what problems I might run into, or when the result may not turn out as good as I imagined. That saves me from wasting time going down the wrong path.}}

In the \textit{Resituate} stage, many valued the feedback generated by the professional artist agent for its constructive insights, while the fan agent helped highlight their distinctive traits and what might be lost when shifting styles. 
For example, P14 shared, \enquote{\textit{It pointed out that my work was missing particle effects, something I usually include. That reminded me how to reintroduce my own style into the concept art so it truly feels like my own.}} Feedback from trending viewers was often considered less insightful, frequently compared to \enquote{\textit{passerby comments}} (P13), though some still saw its value in commercial contexts, as P16 noticed that the artwork has to be rich and memorable enough for general viewers to notice. 

In contrast to AER, the baseline framed style exploration primarily as prompt iteration rather than explicit interpretation or reflection on style. Most artists did not begin by interpreting references (P1, P5, P7, P13, P14, P16, P19), but instead started with simple prompts such as \enquote{\textit{use this style to generate the artwork}} and refined subsequent prompts based on what was missing from the outputs. As a result, artists relied mainly on their own aesthetic judgment to extract usable fragments from the generated images and synthesize them into a coherent direction for exploration.
\subsubsection{Challenges and Opportunities in AER-embedded Workflow}
% \paragraph{Flexibility Across the AER Workflow}
\paragraph{Constraints of a Fixed Workflow}
% The artists showed varying emphasis on different steps of the AER workflow. 
Artists selectively emphasized different steps depending on their goals, preferences, and stage of style exploration.
For some (P2, P3, P11), they stated that \textit{Analyze} alone already provided substantial inspiration. As P2 explained, \enquote{\textit{The analysis is powerful. Once I understand the logic, I don't even need it to generate simulated artworks, because I'm an artist, I can learn from it, and I have the skills to make the art.}} Also, A few (P4, P18) often skipped the \textit{Resituate} stage. As P18 explained, \enquote{\textit{Until I actually draw it myself, it's a completely different matter, so I don't need feedback on a work that isn't mine.}} This suggests the need to enable professional artists a flexible workflow to work with creativity support tools during style exploration.  
% This shows the flexibility of the AER workflow, as artists selectively emphasized different steps depending on their needs, preferences, and stage of style exploration. 

%% \textit{"If I already understand the style elements, I don't even need it to generate simulated artworks. The analysis is powerful. [...] Once I understand the logic, I can learn from it, because I'm an artist, and I have the skills to make the art.”} 

\paragraph{AER Framework Resembles Artists' Daily Practices}
Many participants emphasized that the AER framework felt closely aligned with their everyday practices of style exploration and learning. P1 described its versatility across different stages of professional work: \enquote{\textit{With [AER], I can see myself using it during the ideation phase, when I'm preparing to pitch to clients, or even after completing a project if I want further to explore style variations. It feels much more integrated with my actual [creativity] workflow.}} P13 similarly compared the workflow to the teaching and analysis process: \enquote{\textit{This [AER workflow] is very close to how I teach students, telling them how to look at the atmosphere, composition, and techniques.}} 

Thus, these findings motivated us to explore the long-term impact of AER framework on style exploration for professional artists.

\section{Field Study}
While the controlled experiment examined the immediate effects of AER on artists' agency and reflection, style exploration is rarely a one-session activity. To understand how AER fits into ongoing creative practice, we conducted a two-week field study with four professional digital artists. We asked: \textbf{how do artists engage with the AER framework over time during style exploration?}

%\textbf{how do artists integrate the AER framework into their daily style exploration practice over time?}

\subsection{Participants and Procedure}
We recruited four professional artists who had participated in both our formative and controlled studies: two environment concept artists, P2 (24 years old, 6 years of experience) and P4 (27 years old, 4 years of experience), and two illustrators, P1 (31 years old, 10 years of experience) and P3 (30 years old, 10 years of experience), both of whom specialized in character illustration with rich environments. Each participant received approximately \$220 USD.

For the field study, we deployed AER as a web-based system for two weeks. We instructed participants to use it for about one hour (a session) per day. While daily use was encouraged, it was not strictly enforced. Participants could skip one or two days and make up their usage later, providing flexibility while ensuring sustained engagement with the system. We updated the system to use Flux 2 Pro\footnote{Flux 2 Pro, https://bfl.ai/models/flux-2}, which provided finer stylistic recombination, and added support for replacing the image at any stage of the AER workflow with either a generated or user-uploaded image. We did not assign a fixed task for them to use the system, allowing participants to appropriate any feature of the Analyze, Experiment, and Resituate stages according to their own creative goals.

After each daily session, participants completed a short diary entry describing how they used the system, any notable discoveries, and satisfying experimentation. At the end of the study, we conducted a semi-structured interview with each participant to reflect on their usage over time. The interviews also drew on participants' diary entries, prompting them to revisit and elaborate on specific experiences and patterns documented across the study.

\subsection{Findings}
\begin{figure}[ht]
  \centering
  \includegraphics[width=\linewidth]{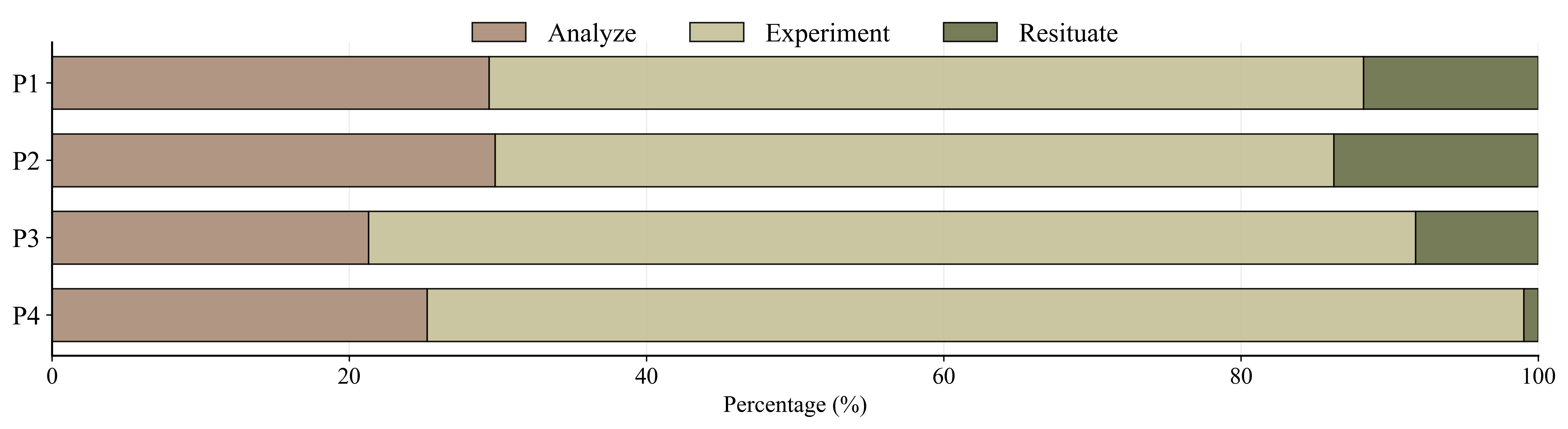}
  \caption{The proportion of usage across the Analyze, Experiment, and Resituate stages during the field study.} 
  \Description{Stacked bar chart showing participants' usage distribution across system stages. Each horizontal bar represents one participant (P1–P4) and sums to 100 percent usage. The three segments correspond to Analyze, Experiment, and Resituate. Across participants, Experiment occupies the largest proportion, Analyze accounts for a moderate share, and Resituate represents the smallest proportion.}
  \label{fig:fieldstudy}
\end{figure}

% In this section, we report findings from the field study that could only be observed through longer-term use. 
Our field study shows how professional artists integrated AER into creative practice over time, how their use evolved, and what this implies for AI tools for artistic exploration. Figure~\ref{fig:fieldstudy} illustrates each participant's proportional usage across three stages, with a mean total usage time of 9.63 hours. On average, participants spent the most time in the \textit{Experiment} stage (64.78 \%), followed by \textit{Analyze} (25.85 \%) and \textit{Resituate} (9.37 \%). The stages were not used in a strictly linear manner; artists moved between them based on their exploration needs.
%  In this part, we focus on the findings that can only be found during long-term study, 這邊的發現非常有趣, 我相信會為未來的AI和專業藝術家的協作上帶來深刻insight
% subsubsection 1
% How AES help artist Reflect on their style exploration and decision process:
% Analyze

\subsubsection{How AER Influenced Reflection during Style Exploration}
\label{sec:AER supported style reflection}
Across participants, AER supported ongoing reflection on stylistic choices, with the three stages contributing differently.

\paragraph{Analyze}
Participants described \textit{Analyze} as helping them externalize tacit stylistic judgments. Rather than introducing entirely new ideas, it helped them revisit previously learned principles, notice overlooked details, and clarify what they valued in a reference image they uploaded to the system. For P3, \textit{Analyze} surfaced aspects they might otherwise ignore in a habitual workflow: \enquote{\textit{My drawing workflow is already quite fixed, and this stage [Analyze] reminded me that there were still other elements I could pay attention to.}} Similarly, P2 felt that: \enquote{\textit{Turning some abstract concepts into text made them more inspectable.}} \textit{Analyze} also helped participants refine their interpretation of references. For example, after reading the stylistic elements in \textit{Analyze}, P1 reflected: \enquote{\textit{In this image, I initially thought what I liked was the linework, especially the hatching, but later I realized it was actually the rendering style—the way shading was expressed through lines.}}

% However, this support was uneven. 
However, some participants felt that the current labels (e.g., Composition) provided by \textit{Analyze} were insufficient. For example, P4 felt the system's analysis often identified visible elements without explaining why they worked in context: \enquote{\textit{while the system could label features such as S-shaped composition or visual guidance, it did not explain how those elements were arranged appropriately to create balance and emphasis within the image.}}

\begin{figure}[ht]
  \centering
  \includegraphics[width=\linewidth]{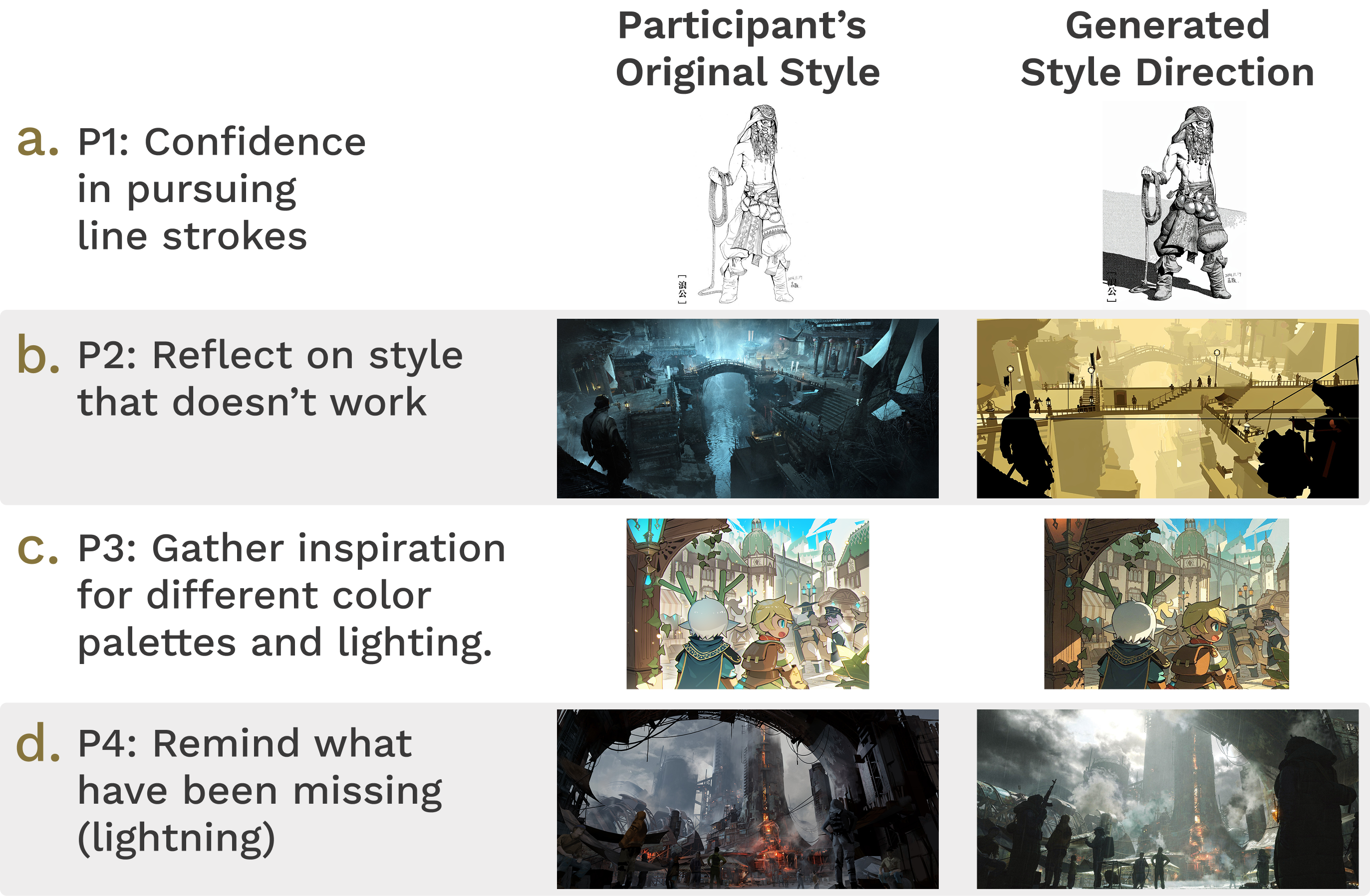}
  \caption{Examples illustrating how the Experiment stage triggers diverse forms of reflection. This stage helps artists (a) gain confidence to pursue unfamiliar stylistic directions, (b) reflect on why certain styles fail, (c) gather inspiration, and (d) identify missing elements in prior work.} 
  \Description{Examples showing how generated variations prompted different types of stylistic reflection. Each row compares an original artwork (left) with a style variation (right). Row (a) shows increased confidence in linework through added texture and shading. Row (b) illustrates reflection on an unsuitable style direction with reduced depth. Row(c) presents gathering inspiration from artwork. Row (d) highlights recognition of missing elements, where enhanced lighting introduces stronger atmosphere.}
  \label{fig:fieldstudy_experiment}
\end{figure}

\paragraph{Experiment}
The \textit{Experiment} stage enabled artists' reflection by making alternative stylistic directions concrete and comparable. By trying different combinations, participants examined how stylistic choices affected mood, composition, and rendering, and in turn reflected on the limits of their prior habits. For example, P4 found that experimenting with lighting revealed what had been missing from their earlier work (Fig. \ref{fig:fieldstudy_experiment}-d): \enquote{\textit{Seeing these results with more dramatic lighting and stronger contrast made me pay more attention to thinking about how to add those elements in the future.}} P2 similarly reflected that repeated experimentation revealed the narrowness of their earlier style: \enquote{\textit{After trying so many results, I realized my past style had been relatively consistent.}} 
This process also increased artists' confidence in pursuing unfamiliar stylistic directions. For example, P1, who had been working on a series of artworks around a theme, reflected that \enquote{\textit{I used to think this series had to stay in one fixed style, but after using the system I realized it could also work very well in a different style.}} He further explained that making stylistic possibilities more concrete gave him more confidence in imagining how a new style might look, which made him \enquote{\textit{willing to experiment more boldly}} in ways he had not found through Pinterest (Fig.~\ref{fig:fieldstudy_experiment}-a).

At the same time, \textit{Experiment} helped participants reflect on why certain stylistic elements failed when applied to their own work. P2 observed that poor results from Experiment often stemmed from mismatches in the underlying spatial and visual logic (Fig. \ref{fig:fieldstudy_experiment}-b): \enquote{\textit{Some compositional and rendering choices only worked well because they were developed together as a whole.}} 
Rather than treating stylistic elements as independently transferable, seeing the unsatisfying results in the \textit{Experiment} stage helped participants recognize conflicts between the reference, their own image, and the ways different stylistic elements interacted.

\paragraph{Resituate}
Participants' experiences with \textit{Resituate} were more mixed. For some participants, it encouraged critical comparison between the simulated feedback from the system and their own. As P2 explained, \enquote{\textit{I wanted to see whether it thought the same as I did,}} and even disagreement with the simulated feedback could \enquote{\textit{prompt reflection on why the system produced a different reading}}. 
However, most participants used Resituate less often because they were unsure about the reference value of its simulated perspectives. P4 questioned \enquote{\textit{from what perspective the AI was looking at the image,}} while P3 said they would trust it more if it were grounded in actual audience data. 
Participant further noted that Resituate tended to assess images as if they were completed works, which limited its usefulness during exploration. P2 noted: \enquote{\textit{rough outputs could still contain promising ideas, and the system's feedback would be more useful if it could identify unrealized potential and suggest how an unfinished work might be developed further.}}

\subsubsection{From Style Replication to Style Exploration}
% Over time, participants' goals shifted from trying to faithfully reproduce a reference style toward using the system more exploratorily. Early on, three participants (P1, P4, and P8) approached the tool with a strong desire for controllable, closely matched results. As P4 recalled, “At first, I hoped it could reproduce the effect 100\%.” However, this goal-oriented use often made the process feel constrained rather than generative. P1 explained, “When the goal is too strong, it becomes hard to enjoy it… I already pre-imagined the target in my head,” so when the system behaved unexpectedly, the mismatch felt frustrating. Over time, participants relaxed this expectation and began treating the system more as a space for experimentation: “Using it more casually, without such a strong goal, often gave me more interesting results”(P1).
% With this more exploratory attitude, participants often found unexpected outputs more inspiring. P1 reflected that “the results that did not fully follow the reference style actually gave me the most surprise.” Likewise, P4 later became more interested in “using surprising combinations to see what interesting things it could produce.” These outputs often suggested new directions rather than better replications; as P4 described, some felt like “the same world, but in a different place or level... creating an unexpected surprise.”

This exploratory shift broadened how participants used references. Rather than copying surface appearance, they began using the system to reinterpret references more structurally. P2 noted that it was \enquote{\textit{not just copying the visual surface, but could also reshape scene elements and proportions,}} making it \enquote{\textit{a more innovative and interesting way for artistic exploration.}} P3 similarly extended this exploratory use beyond artwork references by experimenting with photographs, explaining that \enquote{\textit{carefully staged photographs could produce a stronger mood,}} making them useful not as literal models to copy, but as evocative inputs for exploration.

\begin{figure}[ht]
  \centering
  \includegraphics[width=\linewidth]{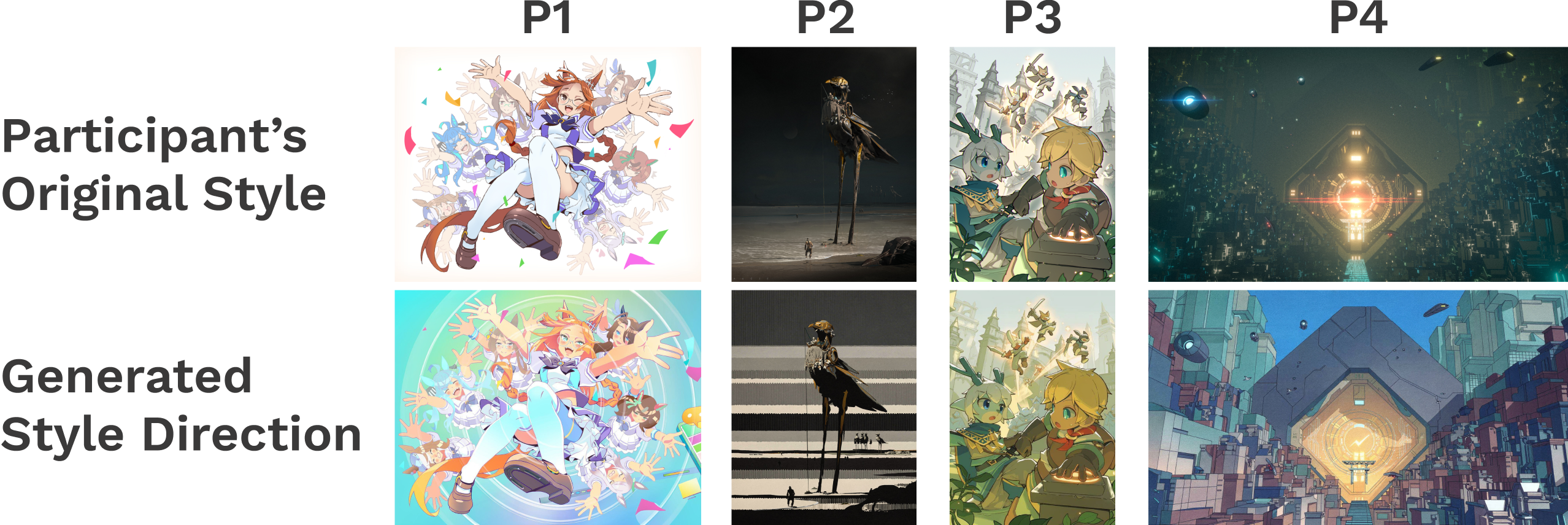}
  \caption{More examples from the Experiment stage.} 
  \Description{Examples showing more artwork results from participants (P1–P4) for the Experiment stage. Each column compares an original artwork (top) with a style variation (down). }
  \label{fig:fieldstudy_add}
\end{figure}

\subsubsection{AER Outputs Became Parts of the Emerging Style}
Participants often treated the generated images as reusable stylistic material: repositories of elements, solutions, and cues that could be carried into later work. For example, P2 described using the system after establishing a rough composition or sketch, then exploring different stylistic directions to extract \enquote{style elements} for further development rather than committing to a single generated result.
Participants also integrated AER into their wider set of reference practices. P4 described combining two images to \enquote{collide} different possibilities: one might provide the desired style or lighting, while another was closer to the intended content, such as a photo or natural scene. This helped them identify reusable details that they might not have discovered through manual exploration alone. P3 similarly suggested that, over time, they might incorporate AER into everyday Pinterest browsing by uploading images for textual analysis, rather than using it for an immediate task.

\section{Discussion}
% Discussion: AER influence
% Generalization
% Discussion: Ethical (Criticize (open new possibilty -> discussion) )
% Disucssion: 很快地看和看慢生圖給artist影響 
% Discussion: Resituate open new design space 
% DIscussion: AES 降低同質性？ -> different effect on Expert and Novice
% Discussion: 怎麼可以generalize 到novice

\subsection{Shaping Stylistic Trajectories with AER}
% 回扣 RQs
\subsubsection{Reclaiming Agency through Process}
Many artists remain hesitant to adopt GenAI tools due to limited interpretability and control~\cite{jiang2023ai, kawakami2024impact, ko2023large}. 
Our findings in Controlled Experiment Study with direct style transfer echo this similar concern. Participants described crafting prompts and then \enquote{\textit{waiting for whatever it [GenAI] wants to show me}} (P13), which highlights the unpredictability inherent in the image generation process. 
AER framework addresses these limitations by embedding interpretability and control into the workflow. \textit{Analyze} breaks reference artworks into stylistic elements grounded in domain knowledge, making analysis explicit and actionable while supporting more precise expression of intent. \textit{Experiment} generates style variations through a two-stage process: first, defining a style direction (what and why), then specifying how to realize it. Unlike prior systems that map text directly to generated visual outputs~\cite{wang2025gentune, evirgen2024text, chen2024autospark}, AER foregrounds the intentionality behind stylistic shifts, an aspect our Formative Study identified as important to style exploration. \textit{Resituate} introduces multi-perspective feedback that broadens creative possibilities and supports reflection and further iteration, consistent with prior work on the value of diverse perspectives for creativity~\cite{han2024teams, shin2025postermate}.
% Together, AER addresses key limitations of direct style transfer, regarding authorship, predictability, and explainability, transforming style exploration from passive generation into an active, human-centered process~\cite{shneiderman2022human, xu2023transitioning, Hois2019How}.

\subsubsection{Supporting Long-Term Reflections}
During the field study, we observed that artists engaged in \textit{reflection-in-action} while interacting with the AER framework, continuously shaping their style exploration. As a process-oriented framework, AER supports ongoing thinking, decision-making, and exploration throughout the creative workflow. 
% These micro-level reflections and adjustments accumulate into a stronger sense of agency, while simultaneously broadening perspectives and encouraging experimentation.
We further observed a shift in artists' mindset: from \textit{style replication} toward \textit{style exploration}. While prior work suggests that excessive reliance on AI can reduce creativity~\cite{doshi2024generative}, our findings indicate that AER instead promotes exploratory thinking and experimentation when artists work with GenAI systems, which may enhance stylistic diversity in the long run. This is particularly important given growing concerns that GenAI systems can lead to aesthetic homogenization~\cite{doshi2024generative, kumar2025human}. Viewed through the lens of reflective creativity support tools' design patterns~\cite{kreminski2021reflective}, AER supports reflection by inferring artists' stylistic intent in \textit{Analyze} and providing interpretive refraction in \textit{Resituate}. 
% Rather than converging on a single \enquote{correct} answer, 
These interactions encourage artists to question assumptions, justify decisions, and reflect more deeply on their stylistic goals.

%Viewed through the lens of reflective CST design patterns~\cite{kreminski2021reflective}, AER supports reflection by both inferring intent and providing interpretive refraction. In Analyze, the system infers stylistic intent by surfacing technical and conceptual characteristics embedded in reference artworks, helping artists articulate preferences that were previously implicit. In Resituate, multiple persona-based interpretations encourage artists to reconsider their work through alternative viewpoints, creating productive friction rather than converging on a single “correct” answer. This deliberate exposure to diverse and sometimes conflicting interpretations slowed decision-making in meaningful ways, prompting artists to question assumptions, justify choices, and reflect more deeply on their stylistic goals.

\subsection{Integrating AER Framework into the AI-assisted Creative Process}
% Flexibility, DDS generalization 更高層的甚麼可以影響到其他professional field, (新手老手)

Artists did not treat AER workflow as strictly linear; instead, they flexibly adapted stages to fit their own practices. In the controlled experiment study, several participants also noted the potential of applying \textit{Analyze} to their own prior artworks as a source of inspiration. This suggests that the AER framework positions artists as the primary drivers of the creative process.

This also highlights an important characteristic of our participants, professional artists, whose expertise is particularly valuable for understanding how GenAI can support creative practice. Their experience enabled them to prioritize stylistically relevant information when co-exploring style with GenAI. 
%As P3 noted in the formative study, \textit{“Professional artists may be even stronger in the GenAI era, as they understand how and why to create.”} 

At the same time, system-driven analysis in the \textit{Analyze} stage raises concerns. While automated analysis can surface stylistic elements that artists might otherwise overlook, it can potentially make humans gradually offload this important analytical ability to AI-assisted creativity-support tools.
% it may also reduce opportunities for personal interpretation, where much creative insight forms. 
Prior work cautions that overreliance on AI can hinder skill development~\cite{macnamara2024does}, highlighting the need to balance automation with opportunities for interpretive reasoning~\cite{tankelevitch2024metacognitive}. We encourage future work to further examine how system-driven decomposition influences the creative process, as well as how GenAI can be designed to support artists with varying levels of expertise, particularly novice artists.

% More broadly, AER condenses the creative cycle artists already practice: 
AER framework reflects artists' actual practices for style exploration, including studying references, experimenting, and seeking feedback. We expect this three-part pattern could be extended to other art domains: For example, in music, musicians analyze influences (analyze), test motifs (experiment), and share drafts for critique (resituate)~\cite{fleith2000creation, heroux2018creative}. Similar structures appear in writing~\cite{driscoll2020genre, shi2021exploring}, product design~\cite{bigelow2016iterating}, and other creative fields. 
% AER opens broader implications for designing GenAI for creativity support. 
% As agentic workflows emerge that translate high-level inputs into polished outputs~\cite{xi2025rise, wang2024llmagent}, AER lens offers a reframing. 
Rather than collapsing reasoning into end-to-end automation, creativity-support systems should be structured around interpretable elements, user-driven recombination, and embedded reflection. Viewed this way, AER-embedded GenAI shifts from an automation tool to an infrastructure that scaffolds exploration and creative growth.
We encourage further research to examine how the AER framework supports style exploration in fields beyond digital arts.

\subsection{Reflection and Future Work}
Unlike prior work that emphasizes efficiency or output quality in AI-assisted creative works, our approach focuses on designing a GenAI workflow that aligns with artists' practice, distinguishing it from reproduction tools in the creative supply chain~\cite{porquet2025copying}. However, our implementation allows users to reference other artists' works and generate images in their style, often without consent~\cite{moayeri2024rethinking, porquet2025copying}. AER may also further encourage such referencing through style transfer models.
Drawing inspiration from other artists is a common practice in the art world. While AER aims to help artists find inspiration effectively through referencing and analyzing, we acknowledge that using GenAI can raise ethical concerns. These include issues related to style mimicry, copyright risks, and misattribution~\cite{shan2023glaze}. 
Prior work highlights how creative practices are supported within broader ecosystems~\cite{almeda2025creativity}. Building on this perspective, we call for future research to explore how GenAI systems can better embed mechanisms for attribution, consent, and credit, ensuring that artistic contributions are recognized and sustained.

Additionally, in our \textit{Analyze} implementation, we use MLLMs to analyze artworks and evaluate their interactional usefulness, as artists can inspect, challenge, and selectively incorporate the outputs during style exploration. However, analyses derived solely from visual features may still produce context-insensitive interpretations. Artistic style is shaped by cultural influences, historical movements, and dialogues with preceding artists~\cite{de1994spirit}. In our formative study, participants expressed a desire to understand the artists, movements, and influences behind artworks. Such provenance is often absent from current GenAI systems. Future AER-based systems could address this limitation by allowing artists to annotate references with cultural, historical, or biographical context and by incorporating retrieval-augmented models grounded in such information. This direction aligns with emerging HCI work on relationship-aware art exploration~\cite{almeda2025artographer}. Future validation could compare generated analyses against expert artist annotations and assess whether the outputs are both plausible and contextually grounded.

Furthermore, while the \textit{Resituate} stage introduces simulated social roles to diversify artists' perspectives, our findings revealed a tension regarding the transparency and focus of AI-generated feedback. Users noted that their lack of understanding of AI evaluation mechanisms and logic led to psychological resistance to feedback (Section \ref{sec:AER supported style reflection}). Future work could explore different methods to explain the relationships between AI-generated feedback and specific stylistic elements. We also acknowledge that including a \enquote{trending audience} persona is a value-laden design choice that should be implemented with caution. It was introduced to reflect pressures artists already reported in our formative study, rather than to prescribe platform visibility as a goal. The three personas (professional artist, trending audience, and fan) were derived from our formative study and represent viewpoints artists in our study commonly encounter in practice. Importantly, \textit{Resituate} is intended to provide perspectives for reflection, not to simulate actual audience reception or replace human peer feedback. Consistent with our findings, artists selectively adopted feedback based on their own goals and judgment; future implementations should provide greater control over persona characteristics. Looking ahead, building on previous research like Proxona ~\cite{choi2025proxona}, future iterations of the AER framework could integrate artists' social media metrics and historical audience data. Grounding AI feedback in authentic data could transform generic suggestions into more trusted and personalized insights. 

\begin{acks}
This work was partly supported by the National Science and Technology Council (NSTC), Taiwan (under NTSC 114-2221-E-002-218-MY3, and 114-2218-E-002-006), National Taiwan University (114L900902 and 115L8909) funded through the Ministry of Education (MOE), Taiwan; and by Google research grant. We also extend our gratitude to all participants and reviewers for their valuable feedback.
\end{acks}

\bibliographystyle{ACM-Reference-Format}

\bibliography{bibliography}

@article{schapiro1994theory,
  title={Theory and philosophy of art: Style, artist, and society},
  author={Schapiro, Meyer},
  year={1994}
}

@article{ross2003style,
  title={Style in art},
  author={Ross, Stephanie},
  journal={The Oxford handbook of aesthetics},
  pages={228--244},
  year={2003},
  publisher={Oxford University Press Oxford}
}

@article{van2015toward,
  title={Toward Discovery of the Artist's Style: Learning to recognize artists by their artworks},
  author={Van Noord, Nanne and Hendriks, Ella and Postma, Eric},
  journal={IEEE Signal Processing Magazine},
  volume={32},
  number={4},
  pages={46--54},
  year={2015},
  publisher={IEEE}
}

@inproceedings{kim2024audience,
  title={Audience amplified: Virtual audiences in asynchronously performed ar theater},
  author={Kim, You-Jin and Sra, Misha and H{\"o}llerer, Tobias},
  booktitle={2024 IEEE International Symposium on Mixed and Augmented Reality (ISMAR)},
  pages={475--484},
  year={2024},
  organization={IEEE}
}

@incollection{gombrich2018expression,
  title={Expression and communication},
  author={Gombrich, Ernest H},
  booktitle={Modern Art And Modernism},
  pages={177--190},
  year={2018},
  publisher={Routledge}
}

@article{fischer1961art,
  title={Art Styles as Cultural Cognitive Maps 1},
  author={Fischer, John L},
  journal={American anthropologist},
  volume={63},
  number={1},
  pages={79--93},
  year={1961},
  publisher={Wiley Online Library}
}

@article{sherman2017art,
  title={What is art good for? The socio-epistemic value of art},
  author={Sherman, Aleksandra and Morrissey, Clair},
  journal={Frontiers in human neuroscience},
  volume={11},
  pages={411},
  year={2017},
  publisher={Frontiers Media SA}
}

@article{kemp2021artistic,
  title={The artistic expression of feeling},
  author={Kemp, Gary},
  journal={Philosophia},
  volume={49},
  number={1},
  pages={315--332},
  year={2021},
  publisher={Springer}
}

@article{gryglewski2020art,
  title={Art as a message realized through various means of artistic expression.},
  author={Gryglewski, Piotr and Ivashko, Yulia and Chernyshev, Denys and Dmytrenko, Andrii and others},
  journal={Art Inquiry},
  number={22},
  year={2020}
}

@book{pandey2024understanding,
  title={Understanding of Visual Arts Theory and Practice.},
  author={Pandey, Rajkumar},
  year={2024},
  publisher={Blue Rose Publishers}
}

@book{panofsky1995three,
  title={Three essays on style},
  author={Panofsky, Erwin and Heckscher, William Sebastian},
  year={1995},
  publisher={Mit Press}
}

@article{hu2018creative,
  title={Creative self-efficacy as moderator of the influence of evaluation on artistic creativity.},
  author={Hu, Weiping and Wang, Xiaojuan and Yi, Lan Yu Xinfa and Runco, Mark A},
  journal={The international Journal of Creativity \& problem solving},
  year={2018},
  publisher={Korean Assn for Thinking Development}
}

@inproceedings{kang2019art,
  title={Art in the age of social media: Interaction behavior analysis of Instagram art accounts},
  author={Kang, Xin and Chen, Wenyin and Kang, Jian},
  booktitle={Informatics},
  volume={6},
  number={4},
  pages={52},
  year={2019},
  organization={MDPI}
}

@article{guo2024emotional,
  title={Emotional expression in artworks and psychological reactions of audiences},
  author={Guo, Siyan and others},
  journal={Journal of Art, Culture and Philosophical Studies},
  volume={1},
  number={3},
  year={2024}
}

@article{loomis1984cognitive,
  title={Cognitive styles as predictors of artistic styles},
  author={Loomis, Mary and Saltz, Eli},
  journal={Journal of Personality},
  volume={52},
  number={1},
  pages={22--35},
  year={1984},
  publisher={Wiley Online Library}
}

@incollection{abuhamdeh2015artistic,
  title={The artistic personality: A systems perspective},
  author={Abuhamdeh, Sami and Csikszentmihalyi, Mihaly},
  booktitle={The Systems Model of Creativity: The Collected Works of Mihaly Csikszentmihalyi},
  pages={227--237},
  year={2015},
  publisher={Springer}
}

@article{gelade2002creative,
  title={Creative style, personality, and artistic endeavor},
  author={Gelade, Garry A},
  journal={Genetic, Social, and General Psychology Monographs},
  volume={128},
  number={3},
  pages={213},
  year={2002},
  publisher={Kirkpatrick Jordon Foundation}
}

@article{annum2014digital,
  title={Digital painting evolution: A multimedia technological platform for expressivity in fine art painting},
  author={Annum, GY},
  journal={Journal of Fine and Studio Art},
  volume={4},
  number={1},
  pages={1--8},
  year={2014},
  publisher={Academic Journals}
}

@inproceedings{porquet2025copying,
  title={Copying style, Extracting value: Illustrators' Perception of AI Style Transfer and its Impact on Creative Labor},
  author={Porquet, Julien and Wang, Sitong and Chilton, Lydia B},
  booktitle={Proceedings of the 2025 CHI Conference on Human Factors in Computing Systems},
  pages={1--16},
  year={2025}
}

@article{epstein2023art,
  title={Art and the science of generative AI},
  author={Epstein, Ziv and Hertzmann, Aaron and Investigators of Human Creativity and Akten, Memo and Farid, Hany and Fjeld, Jessica and Frank, Morgan R and Groh, Matthew and Herman, Laura and Leach, Neil and others},
  journal={Science},
  volume={380},
  number={6650},
  pages={1110--1111},
  year={2023},
  publisher={American Association for the Advancement of Science}
}

@inproceedings{jiang2023ai,
  title={AI Art and its Impact on Artists},
  author={Jiang, Harry H and Brown, Lauren and Cheng, Jessica and Khan, Mehtab and Gupta, Abhishek and Workman, Deja and Hanna, Alex and Flowers, Johnathan and Gebru, Timnit},
  booktitle={Proceedings of the 2023 AAAI/ACM Conference on AI, Ethics, and Society},
  pages={363--374},
  year={2023}
}

@inproceedings{gatys2016image,
  title={Image style transfer using convolutional neural networks},
  author={Gatys, Leon A and Ecker, Alexander S and Bethge, Matthias},
  booktitle={Proceedings of the IEEE conference on computer vision and pattern recognition},
  pages={2414--2423},
  year={2016}
}

@inproceedings{leitch2025unlimited,
  title={Unlimited Editions: Documenting Human Style in AI Art Generation},
  author={Leitch, Alex and Chen, Celia},
  booktitle={Proceedings of the Extended Abstracts of the CHI Conference on Human Factors in Computing Systems},
  pages={1--9},
  year={2025}
}

@inproceedings{seki2013finding,
  title={Finding impressive social content creators: searching for SNS illustrators using feedback on motifs and impressions},
  author={Seki, Yohei and Miyajima, Kiyoto},
  booktitle={Proceedings of the 36th international ACM SIGIR conference on Research and development in information retrieval},
  pages={1041--1044},
  year={2013}
}

@inproceedings{day2016effect,
  title={The effect of customer perceived value on relationship quality between illustrator and fans to recommendation on Facebook},
  author={Day, Min-Yuh and Chuang, Wei-Chun},
  booktitle={2016 IEEE/ACM International Conference on Advances in Social Networks Analysis and Mining (ASONAM)},
  pages={1135--1142},
  year={2016},
  organization={IEEE}
}

@inproceedings{kumar2025human,
author = {Kumar, Harsh and Vincentius, Jonathan and Jordan, Ewan and Anderson, Ashton},
title = {Human Creativity in the Age of LLMs: Randomized Experiments on Divergent and Convergent Thinking},
year = {2025},
isbn = {9798400713941},
publisher = {Association for Computing Machinery},
address = {New York, NY, USA},
url = {https://doi.org/10.1145/3706598.3714198},
doi = {10.1145/3706598.3714198},
booktitle = {Proceedings of the 2025 CHI Conference on Human Factors in Computing Systems},
articleno = {23},
numpages = {18},
location = {
},
series = {CHI '25}
}

@article{doshi2024generative,
  title={Generative AI enhances individual creativity but reduces the collective diversity of novel content},
  author={Doshi, Anil R and Hauser, Oliver P},
  journal={Science Advances},
  volume={10},
  number={28},
  pages={eadn5290},
  year={2024},
  publisher={American Association for the Advancement of Science}
}

@inproceedings{wang2025aideation,
  title={AIdeation: Designing a human-AI collaborative ideation system for concept designers},
  author={Wang, Wen-Fan and Lu, Chien-Ting and Ponsa i Campany{\`a}, Nil and Chen, Bing-Yu and Chen, Mike Y},
  booktitle={Proceedings of the 2025 chi conference on human factors in computing systems},
  pages={1--28},
  year={2025}
}

@inproceedings{wang2025gentune,
  title={GenTune: Toward Traceable Prompts to Improve Controllability of Image Refinement in Environment Design},
  author={Wang, Wen-Fan and Lee, Ting-Ying and Lu, Chien-Ting and Hsu, Che-Wei and Ponsa i Campany{\`a}, Nil and Chen, Yu and Chen, Mike Y and Chen, Bing-Yu},
  booktitle={Proceedings of the 38th Annual ACM Symposium on User Interface Software and Technology},
  pages={1--21},
  year={2025}
}

@inproceedings{cai2023designaid,
  title={DesignAID: Using generative AI and semantic diversity for design inspiration},
  author={Cai, Alice and Rick, Steven R and Heyman, Jennifer L and Zhang, Yanxia and Filipowicz, Alexandre and Hong, Matthew and Klenk, Matt and Malone, Thomas},
  booktitle={Proceedings of The ACM Collective Intelligence Conference},
  pages={1--11},
  year={2023}
}

@inproceedings{zhang2023adding,
  title={Adding conditional control to text-to-image diffusion models},
  author={Zhang, Lvmin and Rao, Anyi and Agrawala, Maneesh},
  booktitle={Proceedings of the IEEE/CVF international conference on computer vision},
  pages={3836--3847},
  year={2023}
}

@inproceedings{wu2024autogen,
  title={Autogen: Enabling next-gen LLM applications via multi-agent conversations},
  author={Wu, Qingyun and Bansal, Gagan and Zhang, Jieyu and Wu, Yiran and Li, Beibin and Zhu, Erkang and Jiang, Li and Zhang, Xiaoyun and Zhang, Shaokun and Liu, Jiale and others},
  booktitle={First Conference on Language Modeling},
  year={2024}
}

@article{macnamara2024does,
  title={Does using artificial intelligence assistance accelerate skill decay and hinder skill development without performers’ awareness?},
  author={Macnamara, Brooke N and Berber, Ibrahim and {\c{C}}avu{\c{s}}o{\u{g}}lu, M Cenk and Krupinski, Elizabeth A and Nallapareddy, Naren and Nelson, Noelle E and Smith, Philip J and Wilson-Delfosse, Amy L and Ray, Soumya},
  journal={Cognitive Research: Principles and Implications},
  volume={9},
  number={1},
  pages={46},
  year={2024},
  publisher={Springer}
}

@article{muller1982rubens,
  title={Rubens's Theory and Practice of the Imitation of Art},
  author={Muller, Jeffrey M},
  journal={The Art Bulletin},
  volume={64},
  number={2},
  pages={229--247},
  year={1982},
  publisher={Taylor \& Francis}
}

@inproceedings{tankelevitch2024metacognitive,
  title={The metacognitive demands and opportunities of generative AI},
  author={Tankelevitch, Lev and Kewenig, Viktor and Simkute, Auste and Scott, Ava Elizabeth and Sarkar, Advait and Sellen, Abigail and Rintel, Sean},
  booktitle={Proceedings of the 2024 CHI Conference on Human Factors in Computing Systems},
  pages={1--24},
  year={2024}
}

@inproceedings{shan2023glaze,
  title={Glaze: Protecting artists from style mimicry by $\{$Text-to-Image$\}$ models},
  author={Shan, Shawn and Cryan, Jenna and Wenger, Emily and Zheng, Haitao and Hanocka, Rana and Zhao, Ben Y},
  booktitle={32nd USENIX Security Symposium (USENIX Security 23)},
  pages={2187--2204},
  year={2023}
}

@article{moayeri2024rethinking,
  title={Rethinking artistic copyright infringements in the era of text-to-image generative models},
  author={Moayeri, Mazda and Basu, Samyadeep and Balasubramanian, Sriram and Kattakinda, Priyatham and Chengini, Atoosa and Brauneis, Robert and Feizi, Soheil},
  journal={arXiv preprint arXiv:2404.08030},
  year={2024}
}

@article{bigelow2016iterating,
  title={Iterating between tools to create and edit visualizations},
  author={Bigelow, Alex and Drucker, Steven and Fisher, Danyel and Meyer, Miriah},
  journal={IEEE transactions on visualization and computer graphics},
  volume={23},
  number={1},
  pages={481--490},
  year={2016},
  publisher={IEEE}
}

@article{shi2021exploring,
  title={Exploring learner engagement with multiple sources of feedback on L2 writing across genres},
  author={Shi, Yali},
  journal={Frontiers in Psychology},
  volume={12},
  pages={758867},
  year={2021},
  publisher={Frontiers Media SA}
}

@article{driscoll2020genre,
  title={Genre knowledge and writing development: Results from the writing transfer project},
  author={Driscoll, Dana Lynn and Paszek, Joseph and Gorzelsky, Gwen and Hayes, Carol L and Jones, Edmund},
  journal={Written Communication},
  volume={37},
  number={1},
  pages={69--103},
  year={2020},
  publisher={SAGE Publications Sage CA: Los Angeles, CA}
}

@article{heroux2018creative,
  title={Creative processes in the shaping of a musical interpretation: a study of nine professional musicians},
  author={H{\'e}roux, Isabelle},
  journal={Frontiers in Psychology},
  volume={9},
  pages={665},
  year={2018},
  publisher={Frontiers Media SA}
}

@article{fleith2000creation,
  title={The creation process of Brazilian musicians},
  author={Fleith, Denise de Souza and RODRIGUES, MARIA ALEXANDRA MILITATO and Viana, Maria Cristina Alves and Cerqueira, Tereza Cristina Siqueira},
  journal={The Journal of Creative Behavior},
  volume={34},
  number={1},
  pages={61--75},
  year={2000},
  publisher={Wiley Online Library}
}

@inproceedings{almeda2025creativity,
  title={Creativity Supportive Ecosystems: A Framework for Understanding Function and Disruption in Online Art Worlds},
  author={Almeda, Shm Garanganao and Kim, Joy O and Hartmann, Bjoern},
  booktitle={Proceedings of the 2025 CHI Conference on Human Factors in Computing Systems},
  pages={1--17},
  year={2025}
}

@inproceedings{ko2023large,
  title={Large-scale text-to-image generation models for visual artists’ creative works},
  author={Ko, Hyung-Kwon and Park, Gwanmo and Jeon, Hyeon and Jo, Jaemin and Kim, Juho and Seo, Jinwook},
  booktitle={Proceedings of the 28th international conference on intelligent user interfaces},
  pages={919--933},
  year={2023}
}

@inproceedings{son2024genquery,
  title={GenQuery: Supporting Expressive Visual Search with Generative Models},
  author={Son, Kihoon and Choi, DaEun and Kim, Tae Soo and Kim, Young-Ho and Kim, Juho},
  booktitle={Proceedings of the CHI Conference on Human Factors in Computing Systems},
  pages={1--19},
  year={2024}
}

@inproceedings{choi2024creativeconnect,
  title={CreativeConnect: Supporting Reference Recombination for Graphic Design Ideation with Generative AI},
  author={Choi, DaEun and Hong, Sumin and Park, Jeongeon and Chung, John Joon Young and Kim, Juho},
  booktitle={Proceedings of the CHI Conference on Human Factors in Computing Systems},
  pages={1--25},
  year={2024}
}

@article{jing2019neural,
  title={Neural style transfer: A review},
  author={Jing, Yongcheng and Yang, Yezhou and Feng, Zunlei and Ye, Jingwen and Yu, Yizhou and Song, Mingli},
  journal={IEEE transactions on visualization and computer graphics},
  volume={26},
  number={11},
  pages={3365--3385},
  year={2019},
  publisher={IEEE}
}

@inproceedings{peng2024designprompt,
  title={DesignPrompt: Using Multimodal Interaction for Design Exploration with Generative AI},
  author={Peng, Xiaohan and Koch, Janin and Mackay, Wendy E},
  booktitle={Proceedings of the 2024 ACM Designing Interactive Systems Conference},
  pages={804--818},
  year={2024}
}

@inproceedings{han2024teams,
  title={When Teams Embrace AI: Human Collaboration Strategies in Generative Prompting in a Creative Design Task},
  author={Han, Yuanning and Qiu, Ziyi and Cheng, Jiale and LC, RAY},
  booktitle={Proceedings of the CHI Conference on Human Factors in Computing Systems},
  pages={1--14},
  year={2024}
}

@inproceedings{kawakami2024impact,
  title={The Impact of Generative AI on Artists},
  author={Kawakami, Reishiro and Venkatagiri, Sukrit},
  booktitle={Proceedings of the 16th Conference on Creativity \& Cognition},
  pages={79--82},
  year={2024}
}

@article{wan2024breaking,
  title={Breaking the Midas Spell: Understanding Progressive Novice-AI Collaboration in Spatial Design},
  author={Wan, Zijun and Tang, Jiawei and Cai, Linghang and Tong, Xin and Liu, Can},
  journal={arXiv preprint arXiv:2410.20124},
  year={2024}
}

@inproceedings{yan2023xcreation,
  title={XCreation: A Graph-based Crossmodal Generative Creativity Support Tool},
  author={Yan, Zihan and Yang, Chunxu and Liang, Qihao and Chen, Xiang'Anthony'},
  booktitle={Proceedings of the 36th Annual ACM Symposium on User Interface Software and Technology},
  pages={1--15},
  year={2023}
}

@inproceedings{chen2024autospark,
  title={AutoSpark: Supporting Automobile Appearance Design Ideation with Kansei Engineering and Generative AI},
  author={Chen, Liuqing and Jing, Qianzhi and Tsang, Yixin and Wang, Qianyi and Liu, Ruocong and Xia, Duowei and Zhou, Yunzhan and Sun, Lingyun},
  booktitle={Proceedings of the 37th Annual ACM Symposium on User Interface Software and Technology},
  pages={1--19},
  year={2024}
}

@inproceedings{lin2025sketchflex,
  title={SketchFlex: Facilitating Spatial-Semantic Coherence in Text-to-Image Generation with Region-Based Sketches},
  author={Lin, Haichuan and Ye, Yilin and Xia, Jiazhi and Zeng, Wei},
  booktitle={Proceedings of the 2025 CHI Conference on Human Factors in Computing Systems},
  pages={1--19},
  year={2025}
}

@inproceedings{evirgen2024text,
author = {Evirgen, Noyan and Wang, Ruolin and Chen, Xiang 'Anthony},
title = {From Text to Pixels: Enhancing User Understanding through Text-to-Image Model Explanations},
year = {2024},
isbn = {9798400705083},
publisher = {Association for Computing Machinery},
address = {New York, NY, USA},
url = {https://doi.org/10.1145/3640543.3645173},
doi = {10.1145/3640543.3645173},
booktitle = {Proceedings of the 29th International Conference on Intelligent User Interfaces},
pages = {74–87},
numpages = {14},
location = {Greenville, SC, USA},
series = {IUI '24}
}

@article{chen2024memovis,
author = {Chen, Chen and Nguyen, Cuong and Groueix, Thibault and Kim, Vladimir G. and Weibel, Nadir},
title = {MemoVis: A GenAI-Powered Tool for Creating Companion Reference Images for 3D Design Feedback},
year = {2024},
issue_date = {October 2024},
publisher = {Association for Computing Machinery},
address = {New York, NY, USA},
volume = {31},
number = {5},
issn = {1073-0516},
url = {https://doi.org/10.1145/3694681},
doi = {10.1145/3694681},
journal = {ACM Trans. Comput.-Hum. Interact.},
month = nov,
articleno = {67},
numpages = {41}
}

@article{tapal2017sense,
  title={The sense of agency scale: A measure of consciously perceived control over one's mind, body, and the immediate environment},
  author={Tapal, Adam and Oren, Ela and Dar, Reuven and Eitam, Baruch},
  journal={Frontiers in psychology},
  volume={8},
  pages={1552},
  year={2017},
  publisher={Frontiers Media SA}
}

@article{chen2001validation,
  title={Validation of a new general self-efficacy scale},
  author={Chen, Gilad and Gully, Stanley M and Eden, Dov},
  journal={Organizational research methods},
  volume={4},
  number={1},
  pages={62--83},
  year={2001},
  publisher={Sage Publications}
}

@inproceedings{bentvelzen2021development,
author = {Bentvelzen, Marit and Niess, Jasmin and Wo\'{z}niak, Miko\l{}aj P. and Wo\'{z}niak, Pawe\l{} W.},
title = {The Development and Validation of the Technology-Supported Reflection Inventory},
year = {2021},
isbn = {9781450380966},
publisher = {Association for Computing Machinery},
address = {New York, NY, USA},
url = {https://doi.org/10.1145/3411764.3445673},
doi = {10.1145/3411764.3445673},
booktitle = {Proceedings of the 2021 CHI Conference on Human Factors in Computing Systems},
articleno = {366},
numpages = {8},
location = {Yokohama, Japan},
series = {CHI '21}
}

@incollection{hertzmann2023image,
  title={Image analogies},
  author={Hertzmann, Aaron and Jacobs, Charles E and Oliver, Nuria and Curless, Brian and Salesin, David H},
  booktitle={Seminal Graphics Papers: Pushing the Boundaries, Volume 2},
  pages={557--570},
  year={2023}
}

@inproceedings{johnson2016perceptual,
  title={Perceptual losses for real-time style transfer and super-resolution},
  author={Johnson, Justin and Alahi, Alexandre and Fei-Fei, Li},
  booktitle={European conference on computer vision},
  pages={694--711},
  year={2016},
  organization={Springer}
}

@inproceedings{li2024styletokenizer,
  title={Styletokenizer: Defining image style by a single instance for controlling diffusion models},
  author={Li, Wen and Fang, Muyuan and Zou, Cheng and Gong, Biao and Zheng, Ruobing and Wang, Meng and Chen, Jingdong and Yang, Ming},
  booktitle={European Conference on Computer Vision},
  pages={110--126},
  year={2024},
  organization={Springer}
}

@article{batifol2025flux,
  title={FLUX. 1 Kontext: Flow Matching for In-Context Image Generation and Editing in Latent Space},
  author={Batifol, Stephen and Blattmann, Andreas and Boesel, Frederic and Consul, Saksham and Diagne, Cyril and Dockhorn, Tim and English, Jack and English, Zion and Esser, Patrick and Kulal, Sumith and others},
  journal={arXiv e-prints},
  pages={arXiv--2506},
  year={2025}
}

@inproceedings{muller2025genaichi,
  title={GenAICHI 2025: Generative AI and HCI at CHI 2025},
  author={Muller, Michael and Chilton, Lydia B and Maher, Mary Lou and Martin, Charles Patrick and Choi, Minsik and Walsh, Greg and Kantosalo, Anna},
  booktitle={Proceedings of the Extended Abstracts of the CHI Conference on Human Factors in Computing Systems},
  pages={1--9},
  year={2025}
}

@inproceedings{hu2020aesthetic,
  title={Aesthetic-aware image style transfer},
  author={Hu, Zhiyuan and Jia, Jia and Liu, Bei and Bu, Yaohua and Fu, Jianlong},
  booktitle={Proceedings of the 28th ACM International Conference on Multimedia},
  pages={3320--3329},
  year={2020}
}

@inproceedings{yen2017listen,
    title={Listen to Others, Listen to Yourself: Combining Feedback Review and Reflection to Improve Iterative Design},
    author={Yen, Yu-Chun Grace and Dow, Steven P. and Gerber, Elizabeth and Bailey, Brian P.},
    booktitle = {Proceedings of the 2017 ACM SIGCHI Conference on Creativity and Cognition},
    pages = {158–170},
    year = {2017}
}

@inproceedings{shin2025postermate,
  title={PosterMate: Audience-driven Collaborative Persona Agents for Poster Design},
  author={Shin, Donghoon and Lee, Daniel and Hsieh, Gary and Chan, Gromit Yeuk-Yin},
  booktitle={Proceedings of the 38th Annual ACM Symposium on User Interface Software and Technology},
  pages={1--20},
  year={2025}
}

@book{gray2016visualizing,
  title={Visualizing research: A guide to the research process in art and design},
  author={Gray, Carole and Malins, Julian},
  year={2016},
  publisher={Routledge}
}

@book{sutton2013painter,
  title={Painter IX Creativity: Digital Artists Handbook},
  author={Sutton, Jeremy},
  year={2013},
  publisher={Routledge}
}

@inproceedings{ritchie2011d,
  title={d. tour: Style-based exploration of design example galleries},
  author={Ritchie, Daniel and Kejriwal, Ankita Arvind and Klemmer, Scott R},
  booktitle={Proceedings of the 24th annual ACM symposium on User interface software and technology},
  pages={165--174},
  year={2011}
}

@inproceedings{anderson2024homogenization,
  title={Homogenization effects of large language models on human creative ideation},
  author={Anderson, Barrett R and Shah, Jash Hemant and Kreminski, Max},
  booktitle={Proceedings of the 16th conference on creativity \& cognition},
  pages={413--425},
  year={2024}
}

@article{lin2026artificial,
  title={Artificial creativity and agency negotiation: Understanding AI-generated visual art from artistic practitioners’ perceptions},
  author={Lin, Cong and Liu, Chenxu},
  journal={Convergence},
  pages={13548565261417315},
  year={2026},
  publisher={SAGE Publications Sage UK: London, England}
}

@article{bernaschina2025visual,
  title={Visual Art and Artificial Intelligence: Tensions between Ethics, Aesthetics, and Authorship in the Algorithmic Age},
  author={Bernaschina, Diego},
  journal={Contemporary Visual Culture and Art},
  volume={1},
  number={1},
  pages={68--78},
  year={2025}
}

@inproceedings{zheng2025artmentor,
title = {ArtMentor: AI-Assisted Evaluation of Artworks to Explore Multimodal Large Language Models Capabilities},
author = {Zheng, Chanjin and Yu, Zengyi and Jiang, Yilin and Zhang, Mingzi and Lu, Xunuo and Jin, Jing and Gao, Liteng},
booktitle = {Proceedings of the 2025 CHI Conference on Human Factors in Computing Systems},
pages = {18},
year = {2025},
publisher = {Association for Computing Machinery}
}

@inproceedings{duan2024generating,
title = {Generating Automatic Feedback on UI Mockups with Large Language Models},
author = {Duan, Peitong and Warner, Jeremy and Li, Yang and Hartmann, Bjoern},
booktitle = {Proceedings of the 2024 CHI Conference on Human Factors in Computing Systems},
pages = {20},
year = {2024},
publisher = {Association for Computing Machinery}
}

@inproceedings{gao2019communication,
title = {Communication Between Artist and Audience: A Case Study of Creation Journey},
author = {Gao, Yajuan and Wu, Jiede and Lee, Sandy and Lin, Rungtai},
booktitle = {Cross-Cultural Design. Culture and Society: 11th International Conference, CCD 2019, Held as Part of the 21st HCI International Conference, HCII 2019, Orlando, FL, USA, July 26–31, 2019, Proceedings, Part II},
pages = {33–44},
year = {2019},
publisher = {Springer-Verlag}
}

@inproceedings{simpson2025infrastructures,
title = {Infrastructures for Inspiration: The Routine of Creative Identity Through Inspiration on the Creative Internet},
author = {Simpson, Ellen and Semaan, Bryan},
booktitle = {Proceedings of the 2025 CHI Conference on Human Factors in Computing Systems},
pages = {16},
year = {2025},
publisher = {Association for Computing Machinery}
}

@inproceedings{chung2022association,
title = {Artist Support Networks: Implications for Future Creativity Support Tools},
author = {Chung, John Joon Young and He, Shiqing and Adar, Eytan},
booktitle = {Proceedings of the 2022 ACM Designing Interactive Systems Conference},
pages = {232–246},
year = {2022},
publisher = {Association for Computing Machinery},
}

@inproceedings{kim2017mosaic,
title = {Mosaic: Designing Online Creative Communities for Sharing Works-in-Progress},
author = {Kim, Joy and Agrawala, Maneesh and Bernstein, Michael S.},
booktitle = {Proceedings of the 2017 ACM Conference on Computer Supported Cooperative Work and Social Computing},
pages = {246–258},
year = {2017},
publisher = {Association for Computing Machinery},
}

@misc{hung2024simtube,
title = {SimTube: Generating Simulated Video Comments through Multimodal AI and User Personas},
author = {Hung, Yu-Kai and Huang, Yun-Chien and Su, Ting-Yu and Lin, Yen-Ting and Cheng, Lung-Pan and Wang, Bryan and Sun, Shao-Hua},
year = {2024}
}

@inproceedings{choi2025proxona,
title = {Proxona: Supporting Creators' Sensemaking and Ideation with LLM-Powered Audience Personas},
author = {Choi, Yoonseo and Kang, Eun Jeong and Choi, Seulgi and Lee, Min Kyung and Kim, Juho},
booktitle = {Proceedings of the 2025 CHI Conference on Human Factors in Computing Systems},
pages = {32},
year = {2025},
publisher = {Association for Computing Machinery}
}

@article{park2025stealing,
title={Stealing Creator's Workflow: A Creator-Inspired Agentic Framework with Iterative Feedback Loop for Improved Scientific Short-form Generation},
author={Park, Jong Inn and Taneja, Maanas and Wang, Qianwen and Kang, Dongyeop},
journal={ArXiv},
year={2025},
volume={abs/2504.18805},
doi={10.48550/arXiv.2504.18805}
}

@inproceedings{benharrak2024writer,
title = {Writer-Defined AI Personas for On-Demand Feedback Generation},
author = {Benharrak, Karim and Zindulka, Tim and Lehmann, Florian and Heuer, Hendrik and Buschek, Daniel},
booktitle = {Proceedings of the 2024 CHI Conference on Human Factors in Computing Systems},
pages = {18},
year = {2024},
publisher = {Association for Computing Machinery}
}

@article{almeda2025artographer,
  title={Artographer: a Curatorial Interface for Art Space Exploration},
  author={Almeda, Shm Garanganao and Chung, John Joon Young and Liu, Sophia and Halperin, Brett and Lu, Yuwen and Hartmann, Bjoern and Kreminski, Max},
  journal={arXiv preprint arXiv:2512.02288},
  year={2025}
}

@article{de1994spirit,
  title={The spirit of abstract expressionism: selected writings},
  author={De Kooning, Elaine and Slivka, Rose and Luyckx, Marjorie},
  journal={(No Title)},
  year={1994}
}

@inproceedings{kreminski2021reflective,
  title={Reflective Creators.},
  author={Kreminski, Max and Mateas, Michael}
}

@article{braun2006using,
  title={Using thematic analysis in psychology},
  author={Braun, Virginia and Clarke, Victoria},
  journal={Qualitative research in psychology},
  volume={3},
  number={2},
  pages={77--101},
  year={2006},
  publisher={Taylor \& Francis}
}

% \newtcolorbox{mytextbox}[1][]{
%   colback=yellow!20,    % Background color
%   colframe=black,       % Frame color
%   fonttitle=\bfseries,  % Title font
%   coltitle=black,       % Title text color
%   enhanced,
%   sharp corners,
%   boxrule=0.5mm,
%   width=\textwidth,     % Box width
%   top=4pt, bottom=4pt,  % Padding
%   left=6pt, right=6pt,  % Padding
%   breakable             % Allows for box to break across pages
% }

% \lstset{
%   basicstyle=\ttfamily,        % Use a monospaced font
%   breaklines=true,             % Automatic line breaking
%   escapeinside={(*@}{@*)},     % Allows escaping to LaTeX between (*@ and @*)
%   columns=fullflexible         % Better alignment
% }

% % Add appendix at the end of the paper
% % \onecolumn
% \appendix

\lstset{
  basicstyle=\ttfamily,        % Use a monospaced font
  breaklines=true,             % Automatic line breaking
  escapeinside={(*@}{@*)},     % Allows escaping to LaTeX between (*@ and @*)
  columns=fullflexible         % Better alignment
}

\appendix
\onecolumn
\clearpage
\small\ttfamily
\section{Appendix A: Participants Demographic}
\label{Appendix_Participant_Demographic}
\begin{table}[h]
\centering
\footnotesize
\setlength{\tabcolsep}{4pt}
\renewcommand{\arraystretch}{1.2}
\begin{tabular}{c c l c c l c c c}
\toprule
\textbf{ID} & \textbf{Age} & \textbf{Identity} & 
\textbf{YoE} &
\makecell{\textbf{AI-assisted}\\\textbf{creations / week}} &
\textbf{Gen-AI tools used} & 
\makecell{\textbf{Formative}\\\textbf{Study}} & 
\makecell{\textbf{Controlled}\\\textbf{Experiment}\\\textbf{Study}} & 
\makecell{\textbf{Field}\\\textbf{Study}}\\
\midrule
1  & 31 & Illustrator & 10  & $\approx$0 & --- & \checkmark & \checkmark & \checkmark \\
2  & 24 & Concept Artist               & 6  & 10       & Midjourney, StableDiffusion, Leonardo, Gemini & \checkmark & \checkmark & \checkmark \\
3  & 30 & Illustrator                  & 10  & $<1$       & ChatGPT(DALL-E), Gemini & \checkmark & \checkmark & \checkmark \\
4  & 27 & Concept Artist               & 4  & 1          & Midjourney, StableDiffusion, FLUX, Gemini & \checkmark & \checkmark & \checkmark \\
5  & 23 & Illustrator                  & 5  & 3--4       & Midjourney; ChatGPT (DALL·E); Leonardo & \checkmark & \checkmark & \\
6  & 33 & Illustrator                  & 8  & $\approx$0 & --- & \checkmark & & \\
7  & 24 & Concept Artist & 6  & 1          & ChatGPT (DALL·E); Gemini & \checkmark & \checkmark & \\
8  & 30 & Illustrator                  & 14 & $\approx$0 & --- & \checkmark & & \\
9  & 38 & Concept Artist               & 12 & $<1$       & Midjourney & \checkmark & \checkmark & \\
10 & 28 & Illustrator                  & 15+& $\approx$0 & --- & \checkmark & & \\
11 & 36 & Illustrator                  & 10 & $\approx$0 & --- &  & \checkmark & \\
12 & 34 & Concept Artist               & 5  & 3          & Midjourney; ChatGPT (DALL·E); Gemini; Lovart &  & \checkmark & \\
13 & 25 & Illustrator                  & 8 & $\approx$0 & --- &  & \checkmark & \\
14 & 24 & Illustrator                  & 5  & 1          & Midjourney; ChatGPT (DALL·E) &  & \checkmark & \\
15 & 30 & Illustrator                  & 7  & $\approx$0 & --- &  & \checkmark & \\
16 & 24 & Illustrator                  & 5  & 2          & ChatGPT (DALL·E) &  & \checkmark & \\
17 & 24 & Illustrator                  & 5  & 2          & Midjourney; Stable Diffusion; ChatGPT (DALL·E) &  & \checkmark & \\
18 & 32 & Illustrator        & 15 & 2--3       & ChatGPT (DALL·E) &  & \checkmark & \\
19 & 23 & Illustrator                  & 5  & $\approx$0 & --- &  & \checkmark & \\
\bottomrule
\end{tabular}
\Description[This table presents 19 participants, aged 23--38 and with up to 15+ years of experience. One column lists the GenAI tools they use (e.g., Midjourney, Stable Diffusion, ChatGPT (DALL·E), Gemini, Leonardo, Lovart), and two columns indicate participation in formative study,  controlled experiment study, and field study.]{This table details demographic and professional information for 19 participants involved in our study of GenAI usage. Each participant is identified by an ID and characterized by age (ranging from 23 to 38), years of experience (YoE; 4 to 15+), and identity (Concept Artist, Illustrator, Animator). The table also lists which GenAI tools each individual uses in their daily work (e.g., Midjourney, Stable Diffusion, ChatGPT (DALL·E), Gemini, Leonardo, Lovart) to show hands-on experience with GenAI tools. In addition, two columns indicate whether participants took part in formative study, controlled experiment study, and field study}
\caption{Demographic Details of Participants Including Age, Identity, GenAI Tools, and Study Participation}
\label{tab:demographics}
\end{table}

\section{Appendix B: Questionnaire for Controlled Experiment Study}
\label{Appendix_Questionnaire}
\small\ttfamily % Set the text to small, monospaced, and left-aligned
\begin{lstlisting}
Style Development Agency Scale 
- I am in full control of the style decisions I make using this approach.  
- I am the author of my creative choices when exploring styles with this approach.  
- My style exploration is guided by my creative intentions when using this approach.  
- The decision of which style directions to pursue is entirely within my hands when using this approach.  
- My style exploration process is planned and directed by me from beginning to end when using this approach.  
- I am completely responsible for the style outcomes that result from my creative decisions using this approach.  

Reference: Sense of Positive Agency (SoPA) - Frontiers in Psychology  

Style Exploration Self-Efficacy (During System Use)
- I will be able to achieve most of the style exploration goals that I have set for myself using this approach.  
- When facing difficult style exploration tasks, I am certain that I will accomplish them with this approach.  
- In general, I think that I can obtain the style insights that are important to me through this approach.  
- I believe I can succeed at exploring any style direction I set my mind to using this approach.  
- I will be able to successfully overcome many style exploration challenges with this approach.  
- I am confident that I can explore different style elements effectively using this approach.  

Reference: Chen, G., Gully, S.M., & Eden, D. (2001). Validation of a New General Self-Efficacy Scale.  

TSRI: Technology-Supported Reflection Inventory
Insight
- Using the approach has been a wake-up call to make changes in my style exploration process.  
- As a result of using the approach, I have changed how I approach style exploration.  
- Using the approach gives me ideas on how to overcome challenges in developing my style.  

Exploration
- I enjoy exploring my style development possibilities with the approach.  
- The approach makes it easy to get an overview of my current style direction.  
- The approach makes it easy to review my past style experiments and progress.  

Comparison
- I will reflect on my style exploration results in the system with other artists.  
- The approach helps me discuss my style exploration process with others.  
- The system makes me think about how my style exploration compares with that of other artists.  

Reference: The Development and Validation of the Technology-Supported Reflection Inventory. CHI 2021.  

\end{lstlisting}

\section{Appendix C: Figures}
\label{Appendix_Figures}
\tiny\ttfamily % Set the text to small, monospaced, and left-aligned

\label{Appendix_Baseline}
\begin{figure}[H]
  \centering
  \includegraphics[width=\textwidth]{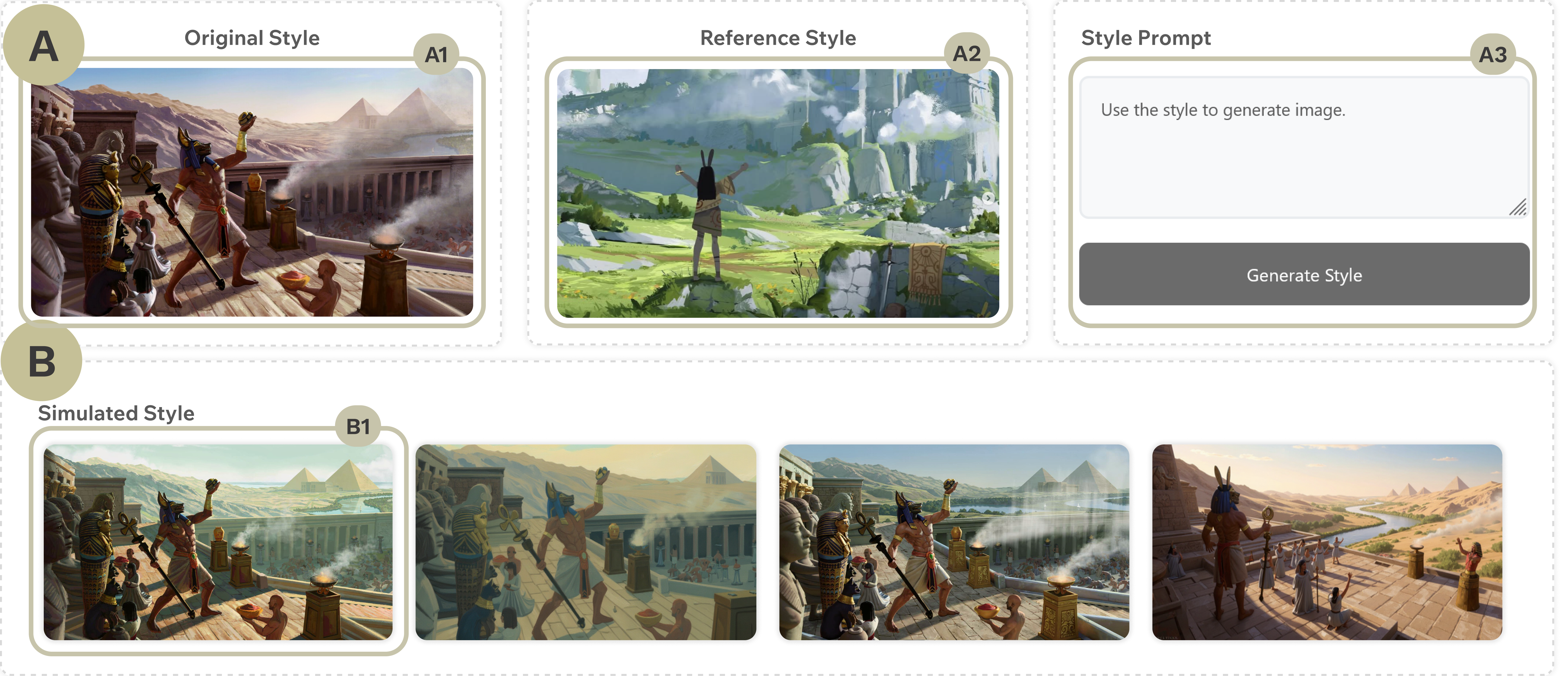}
  \caption{The UI for the non-AER-embedded (direct style transfer) system. (A) Users upload an original personal artwork (A1), a reference artwork (A2) for style reference, and input prompt in the chatbox area (A3). (B) The style simulation section displays 4 style-tweaking artworks merging the content of the user’s artwork with the style of the reference artwork, which users can select (B1).}
  \Description{The A section is divided into three distinct panels. A1 is an artwork section for Original Style and on the right A2 is an artwork section for Referenced Artwork. Then there is a Style Prompt section on the far right, which enables users to type in text commands. Below the chatbox is a button labeled "Generate Style," which initiates the generation process.
The B Section is Simulated Style Output titled "Simulated Style" and serves as the output area for the generated images. The four images are displayed in a horizontal carousel. Each image combines the subject matter and composition of the "Original Artwork" and the style of the "Referenced Artwork". On the far left, one of the generated images is selected as shown by B1.}
  \label{fig:baseline}
\end{figure}

\label{Appendix_workflow}
\begin{figure}[H]
  \centering
  \includegraphics[width=\textwidth]{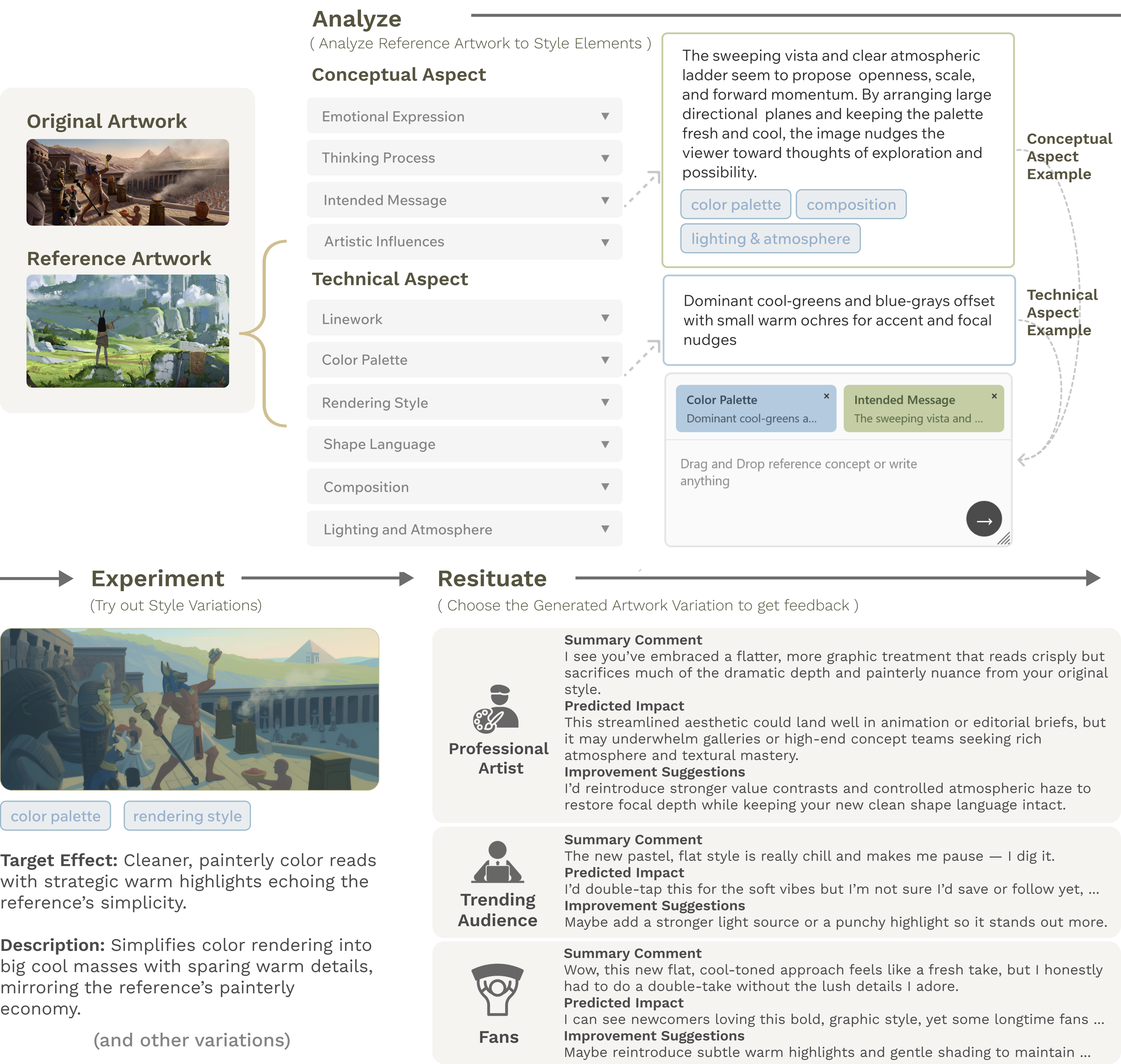}
  \caption{AER workflow. Participants were asked to explore new styles. The AER workflow analyzes the referenced artwork into conceptual aspects, converting the intended message behind the image, and the technical aspects. AER workflow generates image variations based on the decomposed element and produces the text explanations. Then, the workflow simulates three social roles and generates feedback for the selected image variation.}
  \Description{The image displays a workflow diagram illustrating a three-stage process. The diagram is organized linearly from left to right, with headings for each stage: "Analyze," "Experiment," and "Resituate”, with arrows connecting the stages. The Initial part consists of two vertically arranged panels, one is Original Artwork, and the other is Referenced Artwork. The Analyze stage breaks down the Referenced Artwork's style with a series of dropdown menus categorized into "Conceptual Aspect", including Emotional Expression, Thinking Processes, Intended Message, Artistic Influences, and "Technical Aspect" including Linework, Color Palette, Rendering Style, Shape Language, Composition, Lightning and Atmosphere. The diagram also illustrates the detailed descriptions of the dropdown menus. For example, the "Intended Message" dropdown reveals a text box explaining how the "sweeping vista" seems to propose openness, scale, and forward momentum. While the "Color Palette" dropdown is linked to a description of using "Dominant cool-greens and blue-grays." At the bottom of this stage is an interactive box where the user can drag and drop the decomposed stylistic concepts. The next stage is Experiment which visualizes the generated results, the tags of style elements being applied such as "Color Palette," "Rendering Style," "Composition", the Target Effect, and the Description. The final stage is Resituate which provides feedback by simulating responses from different user personas each represented by an icon.  The personas include “Professional Artist”, “Trending Audience”, and “Fans”. The feedback includes a "Summary Comment", "Predicted Impact", and "Improvement Suggestions."}
  \label{fig:system_workflow}
\end{figure}

% \label{Appendix_system_diagram}
% \begin{figure}[H]
%   \centering
%   \includegraphics[width=\textwidth]{figures/Appendix_03_System Diagram.png}
%   \caption{The system diagram for AER-embedded system. Pipeline for Analyze: Once a reference image is provided, Style Elements Extraction MLLM extracted technical and conceptual style elements from the reference artwork.  Pipeline for Experiment: The Style Tweak Direction LLM takes user selected style elements and user prompts as input to construct the style direction. Content Description Extraction MLLM extracts content descriptions from the user's original artwork. Style Prompt LLM take the style direction and the extracted content descriptions to construct the style prompt, which is passed to Flux-Kontext-Pro along with the original and reference artwork to generate artworks
% with descriptions. Pipeline for Resituate: if a user selects a generated artwork, Agent Feedback AutoGen framework takes both the selected artwork and the original artwork to produce agent feedback, and generates simulated feedback.}
%   \Description{The figure shows the three stages of system implementation diagram of the AER-embedded system, labeled Analyze, Experiment and Resituate.}
%   \label{fig:system_diagram}
% \end{figure}

% \input{sections/Supplement}

\end{document}